\documentclass[pra,longbibliography,twocolumn,showpacs,nofootinbib,superscriptaddress,notitlepage]{revtex4-1}

\pdfoutput=1 

\usepackage{dsfont}
\usepackage{amsmath}
\usepackage{amssymb,bm}
\usepackage{amsthm}
\usepackage{mathtools}
\usepackage{url}

\usepackage{array, makecell}
\usepackage{boldline}
\usepackage{enumitem}

\usepackage{color,dsfont}
\usepackage{graphicx}
\usepackage{ragged2e}
\usepackage[colorlinks=true, hyperindex, breaklinks, linkcolor=blue, urlcolor=blue, citecolor=blue]{hyperref} 
\usepackage[normalem]{ulem}
\usepackage[capitalise]{cleveref}
\usepackage{mathrsfs}
\usepackage[caption=false]{subfig}
\usepackage{mathtools}
\usepackage{float}
\usepackage{verbatim}
\usepackage{latexsym}
\usepackage{amsmath}
\usepackage{amssymb}
\usepackage{setspace}
\usepackage{amsfonts}
\usepackage{stmaryrd}
\usepackage{xcolor}
\usepackage{enumitem}
\usepackage{lipsum}

 \newtheorem{definition}{Definition}

\newcommand{\ket}[1]{|#1\rangle}  
 
\newcommand{\codepar}[1]{\ensuremath{[\![#1]\!]}}
\newcommand{\cP}{\mathcal{P}}

\begin{document}
\title{Very low overhead fault-tolerant magic state preparation using redundant ancilla encoding and flag qubits} 
\author{Christopher Chamberland}\email{cchmber@amazon.com}\email{chamber@caltech.edu}
\affiliation{AWS Center for Quantum Computing, Pasadena, CA, 91125, USA}
\affiliation{Institute for Quantum Information and Matter, California Institute of Technology, Pasadena, CA 91125, USA}
\author{Kyungjoo Noh}\email{noh827@gmail.com}
\affiliation{Department of Physics, Yale University, New Haven, CT, 06520, USA}
\begin{abstract}
The overhead cost of performing universal fault-tolerant quantum computation for large scale quantum algorithms is very high. Despite several attempts at alternative schemes, magic state distillation remains one of the most efficient schemes for simulating non-Clifford gates in a fault-tolerant way. However, since magic state distillation circuits are not fault-tolerant, all Clifford operations must be encoded in a large distance code in order to have comparable failure rates with the magic states being distilled. In this work, we introduce a new concept which we call redundant ancilla encoding. The latter combined with flag qubits allows for circuits to both measure stabilizer generators of some code, while also being able to measure global operators to fault-tolerantly prepare magic states, all using nearest neighbor interactions. In particular, we apply such schemes to a planar architecture of the triangular color code family. In addition to our scheme being suitable for experimental implementations, we show that for physical error rates near $10^{-4}$ and under a full circuit-level noise model, our scheme can produce magic states using an order of magnitude fewer qubits and space-time overhead compared to the most competitive magic state distillation schemes. Further, we can take advantage of the fault-tolerance of our circuits to produce magic states with very low logical failure rates using encoded Clifford gates with noise rates comparable to the magic states being injected. Thus, stabilizer operations are not required to be encoded in a very large distance code. Consequently, we believe our scheme to be suitable for implementing fault-tolerant universal quantum computation with hardware currently under development. 
\end{abstract}
\maketitle

\section{Introduction}
\label{section:Introduction}

In order to perform long quantum computations, universal fault-tolerant quantum computers will need to be built with the capability of implementing all gates from a universal gate set with very low logical error rates. Further, the overhead cost for achieving such low error rates will need to be low. Transversal gates are a natural way to implement fault-tolerant gates. Unfortunately, from the Eastin-Knill theorem, given any stabilizer code, there will always be at least one gate in a universal gate set that cannot be implemented using transversal operations at the logical level \cite{EK09}.

Several fault-tolerant methods for implementing gates in a universal gate set have been proposed \cite{KLZ96,PR13,JL14,ADP14,Bombin15,BC15,JB16,YTC16,CJL16,CJL16b,CJ17,ChamberlandMagic,JochymOHomolog19,ZhuUniversal19}. Despite these various proposals, magic state distillation remains a leading candidate in the implementation of a universal fault-tolerant quantum computer \cite{BK05,Reichardt05Mgic,MEK12,BH12,JonesFowlerDist,OC17,Haah2017magicstate,InteSizeStateDist,SubLogOverHaahHastings2018,Haah2018codesprotocols}. Indeed, it had long been believed that implementing magic state distillation was the dominant cost of a universal fault-tolerant quantum computer. While recent results have shown that this is not necessarily the case \cite{Litinski19magicstate}, the cost of performing magic state distillation still remains high. One of the reasons for the high costs of magic state distillation is that the Clifford circuits used to distill the magic states are often not fault-tolerant. Consequently, the Clifford gates must be encoded in some error correcting code (often the surface code) to ensure that these gates have negligible error rates compared to the magic states being injected. 

In Ref.~\cite{ChamberlandMagic}, a fault-tolerant method for directly preparing $\ket{H}$-type magic states was proposed using the Steane code and flag-qubit circuits \cite{YK17,CR17v1,CR17v2,CB17,TCD18Flag,ReichardtFlag18,ChamberlandGKP,ChamberlandPRX,ChaoAnyFlag20,CKYZ20}. For physical error rates $p \gtrsim 10^{-5}$ and with idle qubits failing with error rates 100 times smaller than single-qubit gate error rates, it was shown that fewer qubits were required to prepare $\ket{H}$ states than the best known distillation schemes. Unfortunately, the scheme requires the ability to perform geometrically non-local gates and is scaled by concatenating the Steane code with itself, making it difficult to implement in a scalable way with realistic quantum hardware. 

The Steane code belongs to the family of two-dimensional color codes \cite{Bombin06TopQuantDist,Bombin15,KB15,Kubicathesis}, which are topological codes. In particular, two-dimensional color codes have a nice property that all logical Clifford gates can be implemented using transversal operations. In particular, for a color code with $n$ data qubits, the logical Hadamard gate is simply given by $\overline{H}=H^{\otimes n}$. This features makes color codes particularly well suited for preparing $\ket{H}$-type magic states. Furthermore, recent work introduced a simple and efficient decoding algorithm for color codes \cite{KD19}. Such a decoding scheme was then extended to triangular color codes with boundaries, and a new scalable and efficient decoder which incorporates information from flag qubits was devised, resulting in a competitive threshold of $0.2 \%$ under a full circuit-level depolarizing noise model \cite{CKYZ20}.

In this work we introduce a new fault-tolerant scheme to directly prepare an $\ket{H}$-type magic state encoded in the triangular color code family. We propose an architecture to prepare the $\ket{H}$ state on a two-dimensional planar layout using only nearest neighbor interactions. The preparation scheme is fault-tolerant and achieves full code distance. Such an architecture was made possible not only with flag qubits, but also with a new technique which we call redundant ancilla encoding. In particular, a redundant amount of ancillas are used to measure the color code stabilizers, which turn into flag qubits when fault-tolerantly preparing a GHZ state for measuring the global operator $H^{\otimes n}$. We stress that such a scheme allows one to perform both stabilizer measurements and also measure global operators in a fault-tolerant way without having to change qubit layout. 

Due to the fault-tolerance of our proposed architecture, magic states with very low logical error rates can be directly prepared without the need to use very large distance color codes or surface codes to encode the required Clifford operations. Indeed, we find that the encoded Clifford gates can have logical error rates comparable to the magic states being injected, thus significantly reducing the resource requirements to prepare very high fidelity magic states since the stabilizer operations need not be encoded in a large distance code.  

For physical error rates $p \gtrsim 10^{-4}$ and under a full circuit-level depolarizing noise model, we show that our scheme can be used to prepare magic states with logical error rates comparable to the best known magic state distillation protocols, but with at least an order of magnitude fewer qubits. For instance, for $p = 10^{-4}$, to produce a magic state with a logical error rate of approximately $5 \times 10^{-8}$, our scheme requires only 64 qubits and all Clifford gate operations can be performed at the physical level. Since the magic states can be prepared on a two-dimensional architecture with nearest-neighbor interactions, we believe that our scheme is particularly well suited for quantum hardware currently under development. 

The remainder of the manuscript is structured as follows. In \cref{sec:Prelim} we provide the necessary background information relating to magic states and to the triangular color code family. In \cref{sec:FaultTolHstate} we describe our fault-tolerant magic state preparation protocol using physical Clifford operations. In \cref{sec:Numerics} we provide the necessary details for computing the resource overhead requirements for our magic state preparation scheme and provide detailed numerical results. In \cref{sec:LogicCliffordStatePrep}, we described how our scheme can be used with encoded stabilizer operations, and we provide numerical results for preparing $\ket{H}$ states using such encoded operations. In \cref{sec:Conclusion} we conclude and discuss directions for future work. 

\section{Preliminary material}
\label{sec:Prelim}

In order to make this paper as self-contained as possible, in this section we present the basic preliminary material required to understand the fault-tolerant magic state preparation scheme presented in \cref{sec:FaultTolHstate}. We first introduce $\ket{H}$-type magic states in \cref{HstateSubsec}, followed by the triangular color code family in \cref{subsec:TrainColor} which we use to fault-tolerantly prepare encoded $\ket{H}$ states.

\subsection{$\ket{H}$-type magic states.}
\label{HstateSubsec}
The $n$-qubit Clifford group is defined as 
\begin{align}
\cP_n^{(2)} = \{U : \forall P \in \cP_n^{(1)}, UPU^\dagger \in \cP_n^{(1)}\},
\end{align}
where $ \cP_n^{(1)}$ is the $n$-qubit Pauli group. The Clifford group is generated by
\begin{align}
\cP_n^{(2)} =\langle H, Y\Big(\frac{\pi}{2} \Big), \text{CNOT} \rangle,
\end{align}
where 
\begin{align}
H = \frac{1}{\sqrt{2}} \left( \begin{array}{cc}
                                          1 & 1 \\
                                          1 &-1\\                                                                     
                                          \end{array} \right),
\ \ {\rm and}\ \  
Y\Big( \frac{\pi}{2} \Big) = \frac{1}{\sqrt{2}} \left( \begin{array}{cc}
                                          1 & -1 \\
                                          1 &1\\                                                                     
                                          \end{array} \right).
\label{eq:HOp}
\end{align}
Here $H$ is the Hadamard gate, $Y\Big( \frac{\pi}{2} \Big) = e^{-i \frac{\pi}{4}Y}$ and the CNOT gates acts as
\begin{align}
\text{CNOT} \ket{a} \otimes \ket{b} = \ket{a} \otimes  \ket{a \oplus b},
\label{eq:CNOTgate}                                          
\end{align}
on the computational basis states $\ket{a}$ and $\ket{b}$. The Clifford group, along with the non-Clifford gate\footnote{Unless otherwise specified, whenever we refer to $T$ gates throughout this paper, they will always correspond to the gate given in \cref{eq:TgateDef}.} 
\begin{align}
T = e^{-i\frac{\pi Y}{8}} =  \left( \begin{array}{cc}
                                          \cos{\frac{\pi}{8}} &  -\sin{\frac{\pi}{8}} \\
                                          \sin{\frac{\pi}{8}} & \cos{\frac{\pi}{8}}\\                                                                     
                                          \end{array} \right),
\label{eq:TgateDef}
\end{align}
forms a universal gate set (note that $T = \text{diag}(1,e^{i\pi /4})$ is Clifford equivalent to the $T$ gate defined in \cref{eq:TgateDef}). Hence defining $\mathcal{G} = \langle H, Y\Big( \frac{\pi}{2} \Big), T, \text{CNOT} \rangle$, and given a target fidelity $\epsilon$, a unitary operator $U$ can be approximated with $\mathcal{O}(\log^{c}1/\epsilon)$ gates in $\mathcal{G}$ \cite{Kitaev97,DN06}.

\begin{figure}
	\centering
	\includegraphics[width=0.4\textwidth]{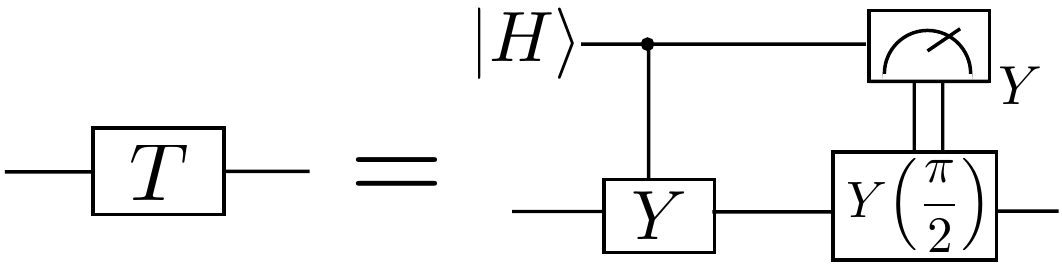}
	\caption{Circuit for simulating a $T$ gate using one copy of an $\ket{H}$ state and stabilizer operations. If the $Y$-basis measurement outcome is $+1$, a $Y\Big( \frac{\pi}{2} \Big)$ gate is applied to the data qubit, otherwise $Y\Big( \frac{\pi}{2} \Big)$ is not applied.}
	\label{fig:TgateCirc}
\end{figure}

A magic state is a state that can be used as a resource state to simulate non-Clifford gates using only stabilizer operations (i.e. Clifford gates, computational basis states and $Z$-basis measurements). Additionally, magic states can also be distilled using only stabilizer operations \cite{BK05}. In this paper we focus entirely on preparing an $\ket{H}$-type magic state \cite{BK05,MEK12}. In particular, an $\ket{H}$ state is given by
\begin{align}
\ket{H} = \cos{\frac{\pi}{8}}\ket{0} + \sin{\frac{\pi}{8}}\ket{1} = T\ket{0},
\end{align} 
which is a $+1$ eigenstate of $H$. Note that $\ket{H}$ is Clifford equivalent to the state 
\begin{align}
\ket{A_{\frac{\pi}{4}}} \equiv \frac{1}{\sqrt{2}}(\ket{0} + e^{i \frac{\pi}{4}}\ket{1}) = e^{i \frac{\pi}{8}}HS^{\dagger}\ket{H},
\end{align}
where $S = \text{diag}(1,i)$ is the phase gate. The state $\ket{A_{\frac{\pi}{4}}}$ can be used to simulate the $T = \text{diag}(1,e^{i\pi /4})$ gate using stabilizer operations. 

\begin{figure}
	\centering
	\subfloat[\label{fig:HadMeasSimple}]{%
		\includegraphics[width=0.2\textwidth]{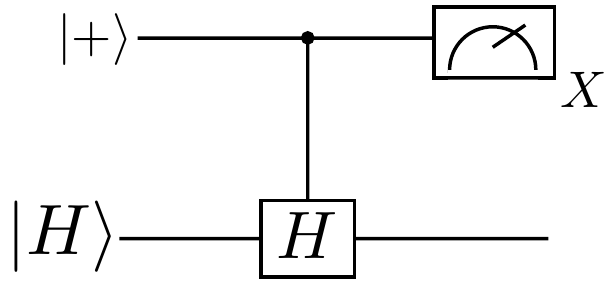}
	}\vfill
	\subfloat[\label{fig:CHdecomp}]{%
		\includegraphics[width=0.3\textwidth]{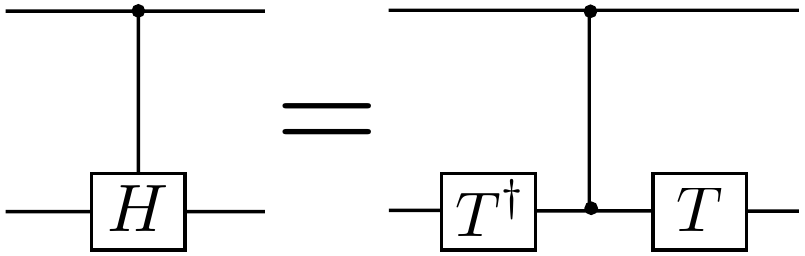}
	}

	\caption{(a) Circuit used to implement a non-destructive measurement of the Hadamard operator. (b) Decomposition of the controlled Hadamard gate in terms of $T$ and $T^{\dagger}$ gates and a controlled-$Z$ gate (which belongs to $\cP_n^{(2)}$). }
\end{figure}

In \cref{fig:TgateCirc}, we provide the circuit used to simulate the $T$ gate in \cref{eq:TgateDef} using one $\ket{H}$ state in addition to stabilizer operations. In many physical implementations however, noisy $\ket{H}$ states are injected into such circuits. To determine if an $\ket{H}$ state is afflicted by an error, one can measure the Hadamard operator using the circuit show in \cref{fig:HadMeasSimple}. Since $HY\ket{H} = -Y\ket{H}$, if $\ket{H}$ is afflicted by a $Y$ error, a $-1$ measurement outcome will be obtained. Further, if $\ket{H}$ is afflicted by an $X$ or $Z$ error, then the measurement outcome are $\pm 1$ at random. If the outcome is $+1$, then no error will be present after the measurement. Note that the controlled-Hadamard gate can be decomposed into products of $T$, $T^{\dagger}$ and controlled-$Z$ gates as shown in \cref{fig:CHdecomp}. 

The goal of this work is to produce encoded $\ket{H}$ states with very low logical failure rates using the fewest possible resources along with an architecture which is suitable for realistic hardware implementations. In particular, we make the Hadamard measurement circuit in \cref{fig:HadMeasSimple} fault-tolerant by using the triangular color code, redundant ancilla encoding, and flag qubits. Along with fault-tolerant implementations of logical Clifford gates, such encoded $\ket{H}$ states could then be used for universal fault-tolerant quantum computation with very low overhead to implement quantum algorithms on near term quantum hardware. 

\subsection{Triangular color code family}
\label{subsec:TrainColor}

\begin{figure}
	\centering
	\includegraphics[width=0.45\textwidth]{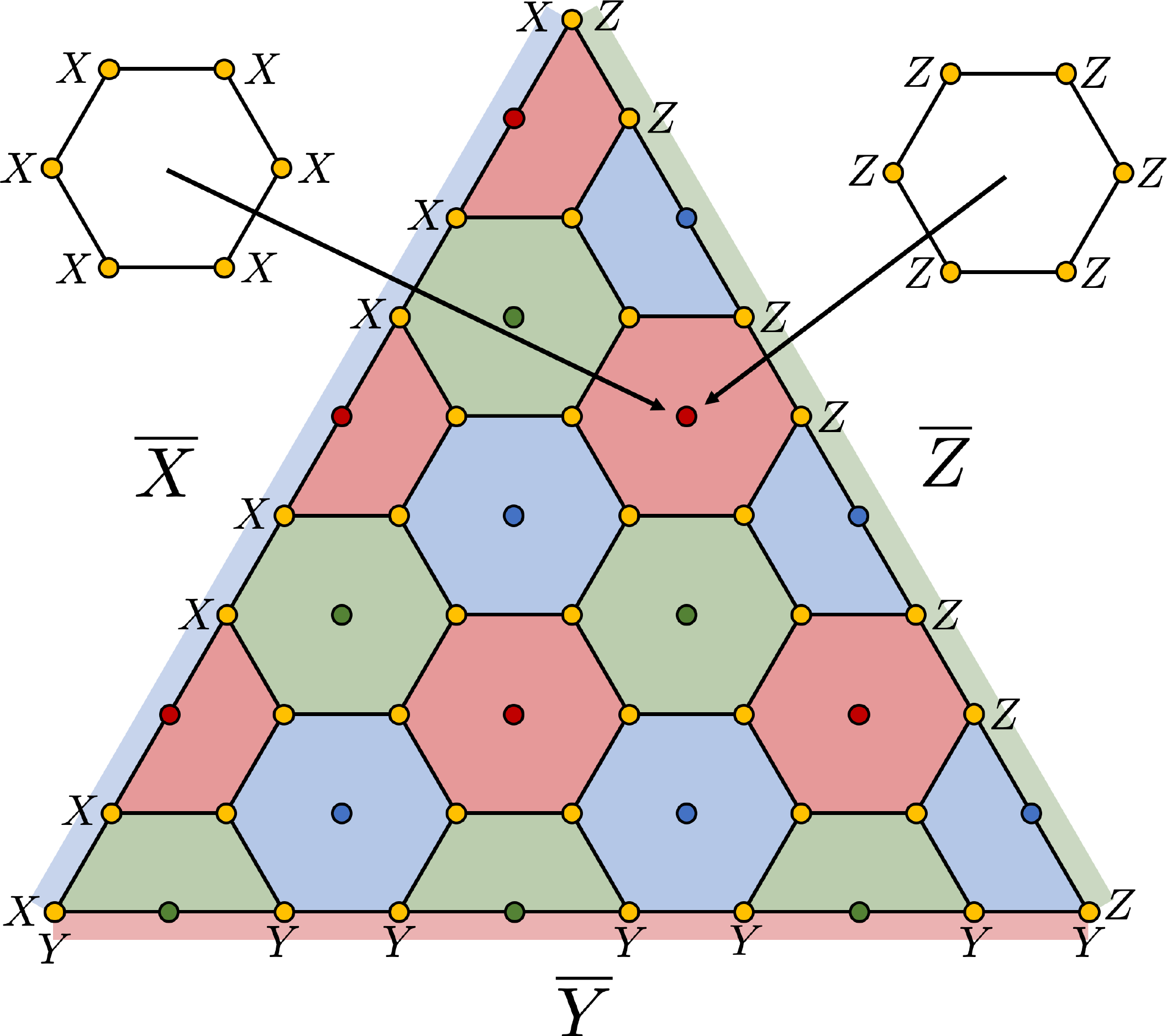}
	\caption{Lattice $\mathcal{L}$ for the implementation of the triangular color code (in this case a distance $d=5$ color code). Each face of $\mathcal{L}$ consists of both $X$ and $Z$-type stabilizer generators which are supported on all qubits belonging to the face. The logical $X$ and $Z$ operators are given by tensor products of all $X$ and all $Z$ operators along a boundary of the triangle. }
	\label{fig:ColorCodeBasic}
\end{figure}

Color codes are topological codes, and thus the data qubits can placed on a lattice where each stabilizer generator can be measured using nearest neighbor interactions. The triangular color code family has code parameters \codepar{n=(3d^2+1)/4,1,d} and is a version of the color code defined on a two-dimensional lattice $\mathcal{L}$ with boundaries. It is a self-dual CSS code with weight-four and weight-six $X$ and $Z$-type stabilizers (see \cref{fig:ColorCodeBasic}).  The lattice $\mathcal{L}$ is 3-colorable, meaning that every face can be colored in red, green or blue with any other face sharing an incident edge having a different color. All vertices of $\mathcal{L}$ (apart from the three corners) are incident to three edges. Further, triangular color codes can implement all logical Clifford gates using transversal operations. In particular, the logical Hadamard operator is simply given by $\overline{H} = H^{\otimes n}$ where $n$ is the number of data qubits. 

In Ref.~\cite{KD19}, an efficient decoder (which we refer to as the \texttt{Lift} decoder) for two-dimensional color codes was provided. In Ref.~\cite{CKYZ20} it was shown how the \texttt{Lift} decoder can be extended to color codes with boundaries. Further, it was shown how the \texttt{Lift} decoder can incorporate measurement outcomes from flag qubits to maintain the effective distance of the code under a full circuit-level noise model (see below). Using such methods, it was found that triangular color codes exhibit a competitive threshold value of $0.2 \%$ under a full circuit-level depolarizing noise model.

It is clear from \cref{HstateSubsec} that the logical Hadamard operator can be measured by transversally applying a controlled Hadamard gate between ancillas and every data qubit of a triangular color code. In \cref{sec:FaultTolHstate}, we focus on providing an architecture for triangular color codes allowing the logical Hadamard operator to be measured in a fault-tolerant way using only nearest neighbor interactions. We then show how an encoded $\ket{H}$-type magic state can be fault-tolerantly prepared with very low qubit and space-time overhead. 

The full circuit-level noise model used throughout all simulations performed in this work with physical stabilizer operations is given as follows:

\begin{enumerate}
        \item With probability $p$, each single-qubit gate location is followed by a Pauli error drawn uniformly and independently from $\{ X,Y,Z \}$.
	\item With probability $p$, each two-qubit gate is followed by a two-qubit Pauli error drawn uniformly and independently from $\{I,X,Y,Z\}^{\otimes 2}\setminus \{I\otimes I\}$.	
	\item With probability $\frac{2p}{3}$, the preparation of the $\ket{0}$ state is replaced by $\ket{1}=X\ket{0}$. Similarly, with probability $\frac{2p}{3}$, the preparation of the $\ket{+}$ state is replaced by $\ket{-}=Z\ket{+}$.
	\item With probability $p$, the preparation of the $\ket{H}$ state is replaced by $P\ket{H}$ where $P$ is a Pauli error drawn uniformly and independently from $\{ X,Y,Z \}$.
	\item With probability $\frac{2p}{3}$, any single qubit measurement has its outcome flipped. 	
	\item With probability $p$, each idle gate location is followed by a Pauli error drawn uniformly and independently from $\{ X,Y,Z \}$.
\end{enumerate}

\section{Fault-tolerant $\ket{H}$ state preparation scheme}
\label{sec:FaultTolHstate}

\begin{figure*}
	\centering
	\includegraphics[width=0.9\textwidth]{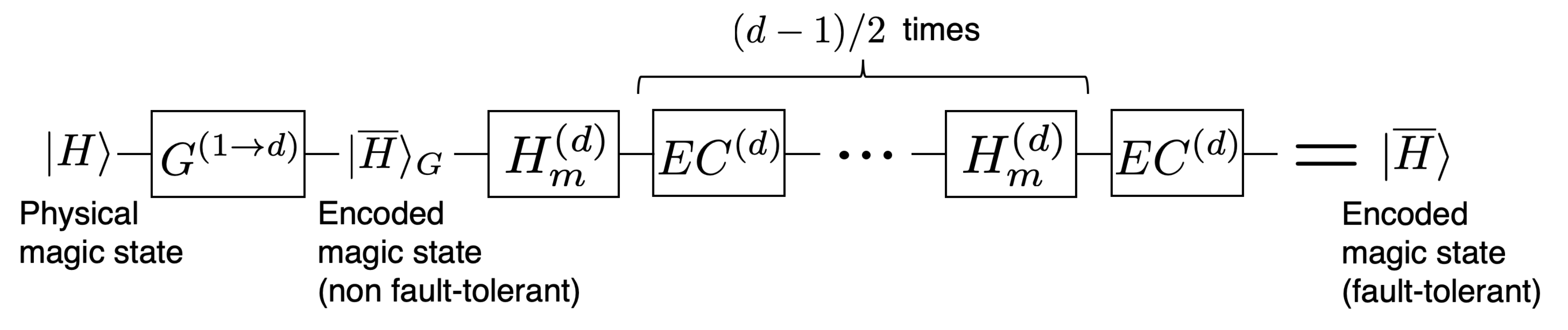}
	\caption{General scheme for the fault-tolerant preparation of an encoded magic state $\ket{\overline{H}}$. First, a physical $\ket{H}$ state is grown to an encoded state of a distance $d$ trianguluar color code, which we label $\ket{\overline{H}}_G$. We represent the circuit which performs the growing operation by the label $G^{(1\rightarrow d)}$. An example for the $G^{(1\rightarrow d)}$ circuit is given in \cref{fig:GrowingPhysical}. Second, a $t$-flag circuit (with $t = (d-1)/2$) for performing a non-destructive measurement of $\overline{H}$ (labelled as $H^{(d)}_{m}$) is applied. In particular, a GHZ ancilla is constructed in a fault-tolerant way to measure $H^{\otimes n}$. An example for one round of the application of the $H^{(d)}_m$ circuit is provided in \cref{fig:MeasureHd5andd7}. Third, error correction for the distance $d$ triangular color code (circuit labelled $EC^{(d)}$) is applied immediately following the $H^{(d)}_m$ circuit (see \cref{fig:ECcircuitFull} for an example when $d=5$). If any flag qubits flag, or the parity of the ancilla measurements in either $H^{(d)}_{m}$ or $EC^{(d)}$ is odd, the protocol is aborted. To guarantee fault-tolerance, each pair of $H^{(d)}_m$ and $EC^{(d)}$ circuits need to be repeated $(d-1)/2$ times (see \cref{appendix:FaultFreeHadMeas}).}
	\label{fig:GeneralHstatePrepScheme}
\end{figure*}

In order for an $\ket{H}$-type magic state to be useful for performing long quantum computations, it is important to be able to prepare such states with very high fidelity. Depending on size and duration of a quantum algorithm, the desired probability that an $\ket{H}$ state is afflicted by an error ranges from $10^{-7}$ to less than $10^{-15}$ \cite{FMMC12,WiebeTgate1,GidneyLowOver19}. Efficient magic state distillation protocols have been devised to prepare such states encoded in an error correcting code with very low error rates \cite{BH12,InteSizeStateDist,Haah2017magicstate,Haah2018codesprotocols,SubLogOverHaahHastings2018,GidneyLowOver19,Litinski19magicstate}. However, as mentioned in \cref{section:Introduction}, magic state distillation circuits are typically not fault-tolerant. Therefore, under a full circuit-level noise model (and since two qubit gates are often the noisiest component of a quantum device), each Clifford operation must be encoded in some large distance error correcting code in order for Clifford gate errors to be negligible. 

We now present a fault-tolerant method for preparing an $\ket{\overline{H}}$ state encoded in a triangular color code using only nearest neighbor interactions\footnote{In what follows, a bar above the $\ket{H}$ symbol always correspond to an encoded $\ket{H}$ state.}. Our method makes use of flag qubits, in addition to a technique which we refer to as redundant ancilla encoding. Since our scheme is fault-tolerant, it will be shown that high fidelity magic states can be obtained using physical Clifford operations in the presence of the full circuit-level depolarizing noise model described in \cref{subsec:TrainColor}. For even higher fidelity magic states, we show in \cref{sec:LogicCliffordStatePrep} that when applying our scheme with logical Clifford operations, such gates can have failure rates which are commensurate with the injected magic states used to implemented $T$  gates (see \cref{fig:TgateCirc,fig:CHdecomp}). Hence only small to intermediate sized codes are necessary to encode the Clifford operations. 

Flag qubits are ancilla qubits used to detect and identify high weight errors arising from a small number of faults between gates which entangle the encoded data with ancillary systems used to perform the necessary measurements for error correction \cite{CR17v1,CB17}. When the measurement outcome of a flag qubit is non-trivial (i.e. $-1$ instead of $+1)$, we say that the flag qubit flagged. We now provide an important definition which is an extension of a definition first introduced in Ref.~\cite{CB17}:
\begin{definition}{\underline{t-flag circuit}}

A circuit $C(U)$ which, when fault-free, implements a projective measurement of a weight-$w$ operator $U \in \cP_n^{(2)}$ without flagging is a $t$-flag circuit if the following holds: For any set of $v$ faults at up to $t$ locations in $C(U)$ resulting in an error $E$ with $\text{min}(\text{wt}(E),\text{wt}(E U)) > v$, the circuit flags. 
\label{Def:tFlaggedCircuitDef}
\end{definition}

In other words, a $t$-flag circuit guarantees that at least one flag qubit flags whenever there are $v \le t$ faults resulting in a data qubit error of weight greater than $v$.

Our scheme begins by growing a physical $\ket{H}$ state into a logical $\ket{\overline{H}}$ state encoded in the distance $d$ triangular color code. When growing a physical $\ket{H}$ state into an encoded $\ket{\overline{H}}$ state, $v < (d-1)/2$ faults can result in an output state of the form $E'\overline{E}\ket{\overline{H}}$ where $E'$ is a detectable error by the color code (i.e. $s(E') \neq \textbf{0}$ where $s(E)$ is the error syndrome of $E$) and $\overline{E}$ is a logical error of the color code. The preparation of $\ket{\overline{H}}$ can be made fault-tolerant by performing an encoded version of the non-destructive Hadamard measurement (the circuit in \cref{fig:HadMeasSimple}) to detect $\overline{E}$ followed by rounds of error correction (EC) to detect the error $E'$ \cite{AGP06,ChamberlandMagic}. The underlying scheme that we use to perform such operations is illustrated in \cref{fig:GeneralHstatePrepScheme}. The scheme consists of three parts. In the first step, we grow a physical $\ket{H}$ state to an encoded $\ket{\overline{H}}_G$ state in a distance $d$ triangular color code, where the circuit which performs the growing operation is labelled $G^{(1\rightarrow d)}$. Note that the state $\ket{\overline{H}}_G$ is obtained using non-fault-tolerant methods. Next, we use $t$-flag circuits (with $t = (d-1)/2$) to perform a non-destructive measurement of $\overline{H}$. The circuit implementing this operations is labelled $H^{(d)}_{m}$. In particular, a GHZ state is constructed in such a way that the $H^{\otimes n}$ measurement is fault-tolerant. Lastly, error correction is performed to detect the errors $E'$ using  $t = (d-1)/2$-flag circuits (labelled $EC^{(d)}$). If any of the flag qubits flag, or the parity of any ancilla measurements in the circuits $H^{(d)}_{m}$ or $EC^{(d)}$ is odd, the protocol is aborted and starts anew.

As shown in \cref{appendix:FaultFreeHadMeas}, in order to guarantee fault-tolerance, both the $H^{(d)}_m$ and $EC^{(d)}$ circuits need to be applied one after the other, and each pair needs to be repeated $(d-1)/2$ times. Further, the circuits used for the $H^{(d)}_m$ and $EC^{(d)}$ operations need to be $t$-flag circuits (with $t = (d-1)/2$) to prevent errors from spreading to uncorrectable errors. In \cref{subsec:TflagHadConstruct,subsec:ECwithRedundantAncilla}, such circuits are provided using only nearest neighbor interactions for $d \in \{3,5,7 \}$. A key idea which allows us to use the same two-dimensional qubit layout and nearest-neighbor interactions to perform \textbf{both} the non-destructive Hadamard measurement and the stabilizer measurements of the triangular color code is the use of redundant ancilla encoding and flag qubits (see \cref{subsec:ECwithRedundantAncilla}).

Note that although our scheme applies for code distances $d \le 7$, in \cref{sec:Numerics} we show that encoded $\ket{\overline{H}}$ states can be prepared with similar logical failure rates but with orders of magnitude fewer qubits (for physical error rates $p \gtrsim 10^{-4}$) compared to some of the state of the art magic state distillation schemes (such as those in Ref.~\cite{Litinski19magicstate}).

\subsection{Growing a physical $\ket{H}$ state to an encoded $\ket{\overline{H}}$ state in a distance $d$ triangular color code}
\label{subsec:GrowingPart}

\begin{figure*}
	\centering
	(a)\includegraphics[width=0.45\textwidth]{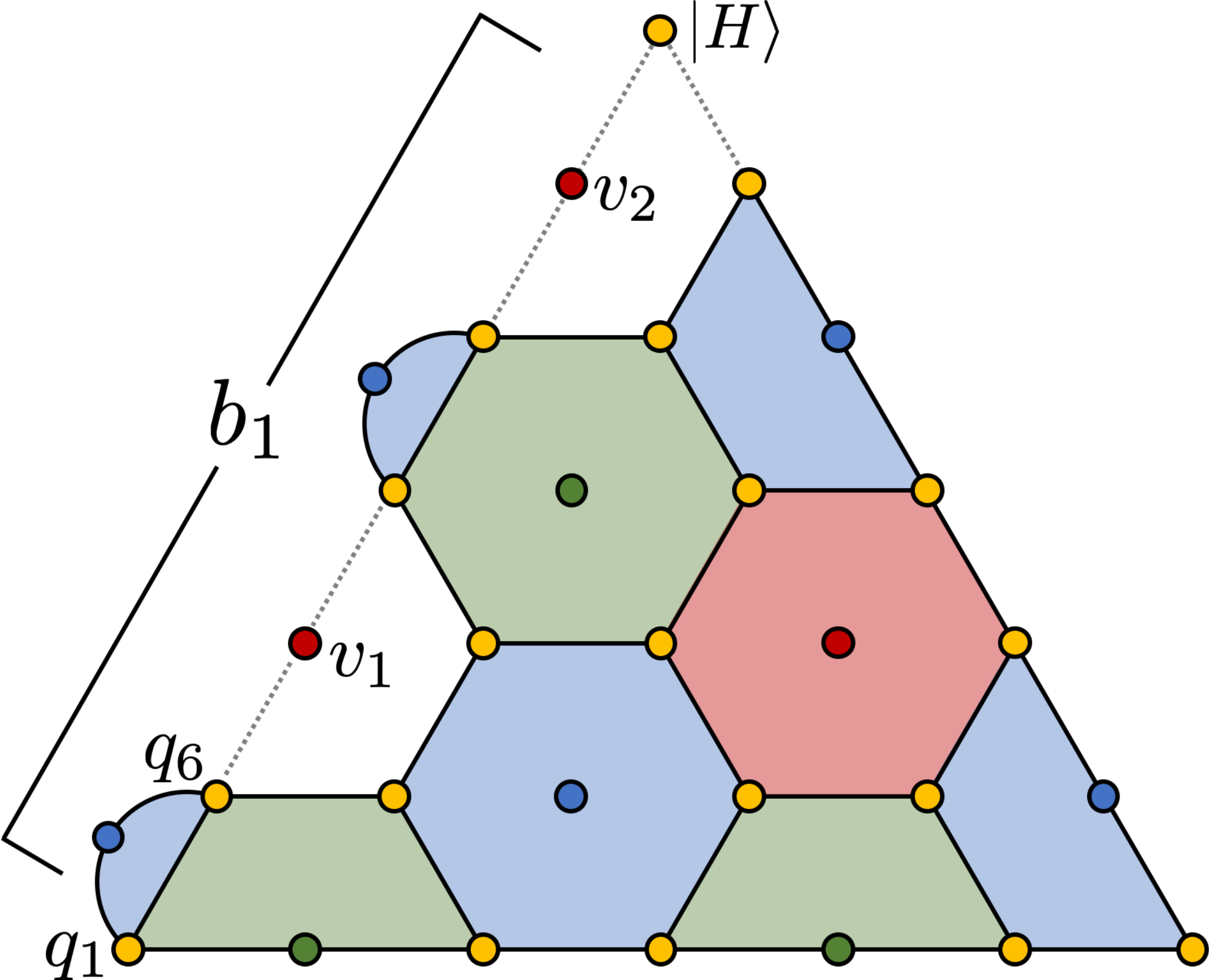}\hspace*{15mm}
         (b)\includegraphics[width=0.2\textwidth]{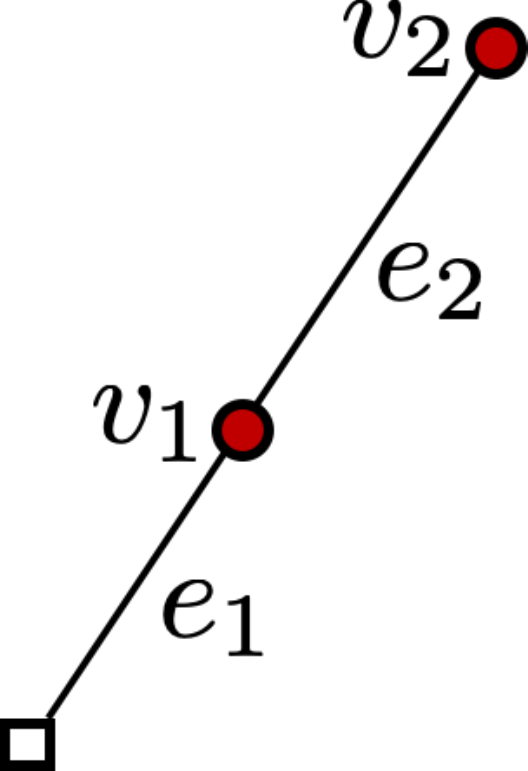}
	\caption{(a) Circuit $G^{(1\rightarrow 5)}$ for growing a physical $\ket{H}$ state to a distance $d=5$ color code. First, a stabilizer state $\ket{S_t}$ is prepared with stabilizers given by the red, green and blue plaquettes. The encoded $\ket{\overline{H}}$ state is then obtained by measuring both weight-four $X$ followed by $Z$-type operators (represented by white plaquettes) along the boundary of the triangle. Such measurements are random since the weight-four operators don't commute with the weight-two stabilizers of $\ket{S_t}$. (b) Matching graph ($G^{(5)}_{1x}$ and $G^{(5)}_{1z}$) used to implement weight-two corrections arising from $-1$ measurement outcomes for generators of $\mathcal{S}_{b_1}$. As an example, if the vertex $v_1$ is highlighted, after implementing MWPM, the edge $e_1$ would be selected resulting in the correction $Z_{q_1}Z_{q_6}$.}
	\label{fig:GrowingPhysical}
\end{figure*}

An illustration of the circuit $G^{(1\rightarrow 5)}$ for growing a physical $\ket{H}$ state to a logical $\ket{\overline{H}}$ state encoded in the $d=5$ triangular color code is given in \cref{fig:GrowingPhysical}a. The preparation of $\ket{\overline{H}}$ in a general distance $d$ triangular color code can be done as follows. First, one prepares a stabilizer state $\ket{S_t}$ (which encodes no logical qubits) that is stabilized by all elements in $\mathcal{S}_{\text{st}} = \mathcal{S}_{w_2} \cup (\mathcal{S}_{\text{color}} \setminus \mathcal{S}_{b_1})$ where $\mathcal{S}_{\text{color}}$ is the stabilizer group of a distance $d$ triangular color code, $\mathcal{S}_{b_1}$ is generated by the $X$ and $Z$-type weight-four operators (white plaquettes) along the boundary $b_1$ of the triangular color code and $\mathcal{S}_{w_2}$ is generated by the weight-two $X$ and $Z$-type operators along the boundary $b_1$ (see \cref{fig:GrowingPhysical} for the case where $d=5$). The qubit which is in the support of $\text{supp}(\mathcal{S}_{\text{color}}) \setminus  \text{supp}(\mathcal{S}_{\text{st}})$ is prepared in the physical $\ket{H}$ state. Note that in the above construction, any of the three boundaries of $\mathcal{L}$ can be chosen. We chose $b_1$ for convention. 

After preparing $\ket{S_t}$, the $X$ and $Z$-type generators of $\mathcal{S}_{b_1}$ are measured along the boundary $b_1$. Since the generators of $\mathcal{S}_{b_1}$ don't commute with the weight-two generators of $\mathcal{S}_{w_2}$, the measurement outcomes of each generator in $\mathcal{S}_{b_1}$ will be $\pm 1$ at random. If a $-1$ outcome is obtained, a Pauli frame update \cite{Knill05,Barbara15,DA07,CIP17} needs to be applied to obtain the correct encoded state. In order to perform the correct Pauli frame update based on the random measurement outcomes, we define two one-dimensional graphs $G^{(d)}_{1x}$ and $G^{(d)}_{1z}$. The vertices of the graph $G^{(d)}_{1x}$ contain the random measurement outcomes of the $X$-type generators in $\mathcal{S}_{b_1}$. Each edge corresponds to two qubits which have support on one of the weight-two operators of $\mathcal{S}_{w_2}$. Similarly, the vertices of the graph $G^{(d)}_{1z}$ encodes the random measurement outcomes of the $Z$-type generators in $\mathcal{S}_{b_1}$. An example is provided in \cref{fig:GrowingPhysical} for the $d=5$ triangular color code. Given the set of highlighted vertices of $G^{(d)}_{1x}$ and $G^{(d)}_{1z}$, Minimum-Weight-Perfect-Matching (MWPM) \cite{Edmonds65} is applied on both graphs. Each highlighted edge involves performing a weight-two $X$ or $Z$-type Pauli frame update. For instance, for the graph $G^{(5)}_{1x}$ in \cref{fig:GrowingPhysical}b, if the edge $e_1$ is selected during MWPM, the correction $Z_{q_1}Z_{q_6}$ is applied to the data. Note that the growing scheme is not fault-tolerant, and thus there is no need to repeat the measurements described above. More details on the implementation of the circuits $G^{(1\rightarrow d)}$ are provided in \cref{appendix:NonFaultTolerantPrepScheme}.

\subsection{Performing error correction ($EC^{(d)}$) using redundant ancilla encoding}
\label{subsec:ECwithRedundantAncilla}

\begin{figure*}
	\centering
	\includegraphics[width=0.7\textwidth]{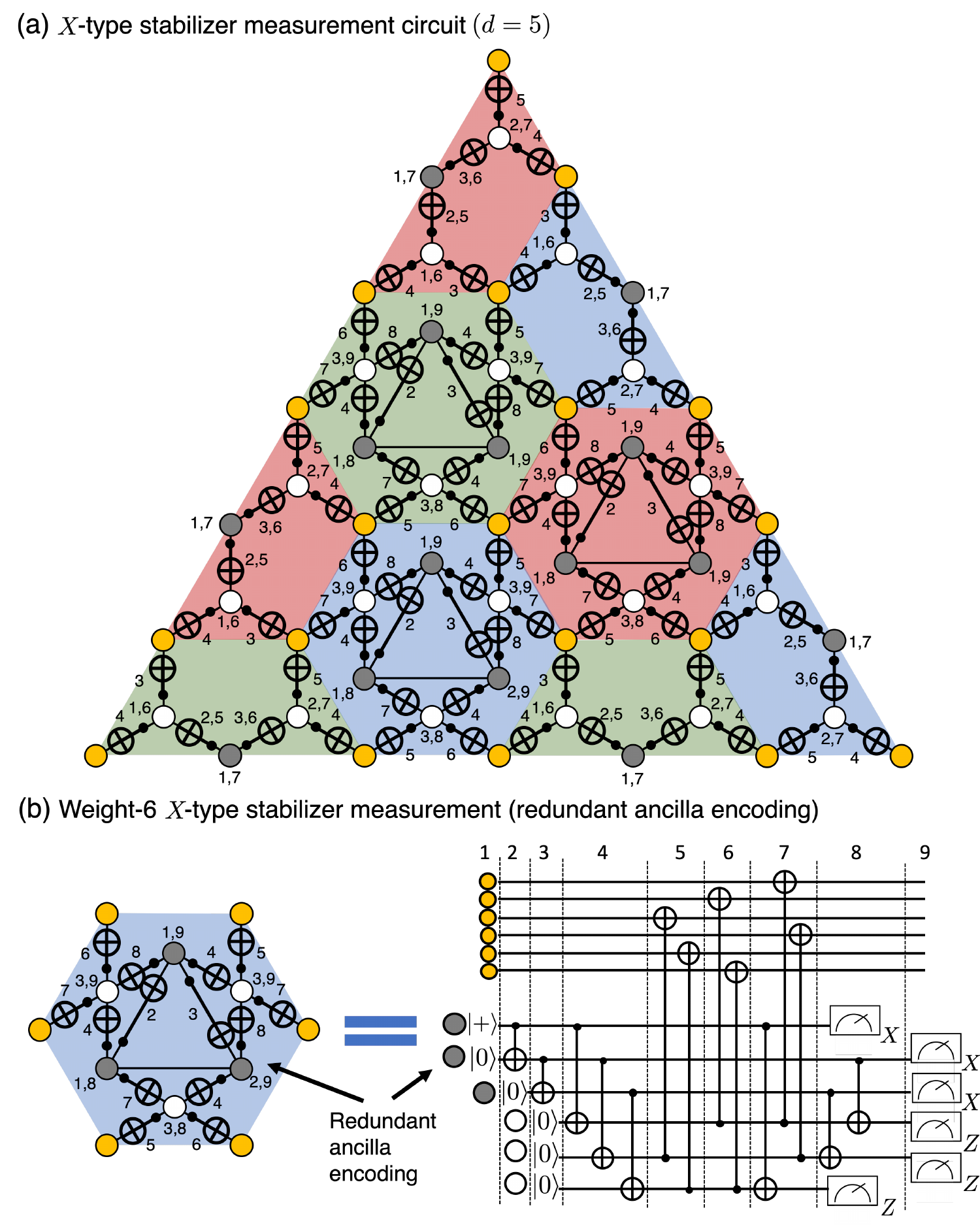}
	\caption{(a) $EC^{(5)}$ circuit for measuring the stabilizers of the $d=5$ color code. The grey circles correspond to the ancilla qubits used to measure the parity of the stabilizers, whereas the white circles correspond to the flag qubits. The CNOT gate scheduling which minimizes the total number of time steps for measuring the $X$ and $Z$-stabilizers is also provided. One round of $X$ or $Z$-type stabilizer measurements requires a total of nine time steps. (b) Circuit for measuring a weight-six $X$ stabilizer of the triangular color code. A redundant amount of ancillas are used since they become flag qubits when measuring $H^{\otimes n}$.}
	\label{fig:ECcircuitFull}
\end{figure*}

In this section we describe the error correction circuits $EC^{(d)}$ used to detect errors at the output of the $H^{(d)}_{m}$ circuits. In particular, to guarantee fault-tolerance (see \cref{Def:FaultTolerantPrep} in \cref{appendix:FaultFreeHadMeas} for our definition of fault-tolerance) of our scheme, we require that each $EC^{(d)}$ circuit is at least a one-flag circuit. For instance, if two faults in two separate $Z$-type stabilizers both result in a weight-two data qubit error without any flag qubits flagging, such an error might not be correctable by a $d=5$ triangular color code. Hence such circuits certainly don't satisfy \cref{Def:FaultTolerantPrep}. Note that in Ref.~\cite{CKYZ20}, it was proved that only one flag circuits are required when performing error correction with the triangular color code. The need for only one-flag circuits has to do with the fact that if two-faults occur during a weight-six stabilizer measurement resulting in a weight-three data qubit error, such an error cannot have full support along a minimum-weight logical operator of the triangular color code.

In \cref{fig:ECcircuitFull}a we illustrate the full $EC^{(5)}$ circuit used to measure the stabilizers of a triangular color code along with the CNOT gate scheduling which minimizes the circuit depth for the given qubit layout. The weight-four stabilizers are identical to the ones used in Ref.~\cite{CKYZ20}. However, an important difference can be observed for the weight-six stabilizers, which in addition to using three flag qubits, also uses three ancilla qubits. If an error anti-commutes with a weight-six stabilizer, the measurement outcomes of the three ancillas will have odd parity, otherwise it will have even parity\footnote{This is similar to error correction circuits used for Shor error correction \cite{Shor96,DS96}. However an important difference is that ancilla verification (i.e. using several ancillas to measure pairs of qubits) is not necessary since the flag qubits ensure that the circuit in \cref{fig:ECcircuitFull}b is a two-flag circuit.}. The circuit for measuring a weight-six $X$-type stabilizer which respects the CNOT scheduling is given in \cref{fig:ECcircuitFull}b. By performing an exhaustive numerical search, we verified that both weight-four and weight-six circuits are two-flag circuits. Since the weights of the stabilizer generators are independent of $d$, having two-flag circuits is sufficient to ensure that the $EC^{(d)}$ circuits are implemented fault-tolerantly \cite{CB17}. Lastly, each weight-four and and weight-six plaquette in a general $EC^{(d)}$ circuit has the same qubit layout and gate connectivity as those of \cref{fig:ECcircuitFull}a.

\begin{figure*}
	\centering
	\includegraphics[width=0.9\textwidth]{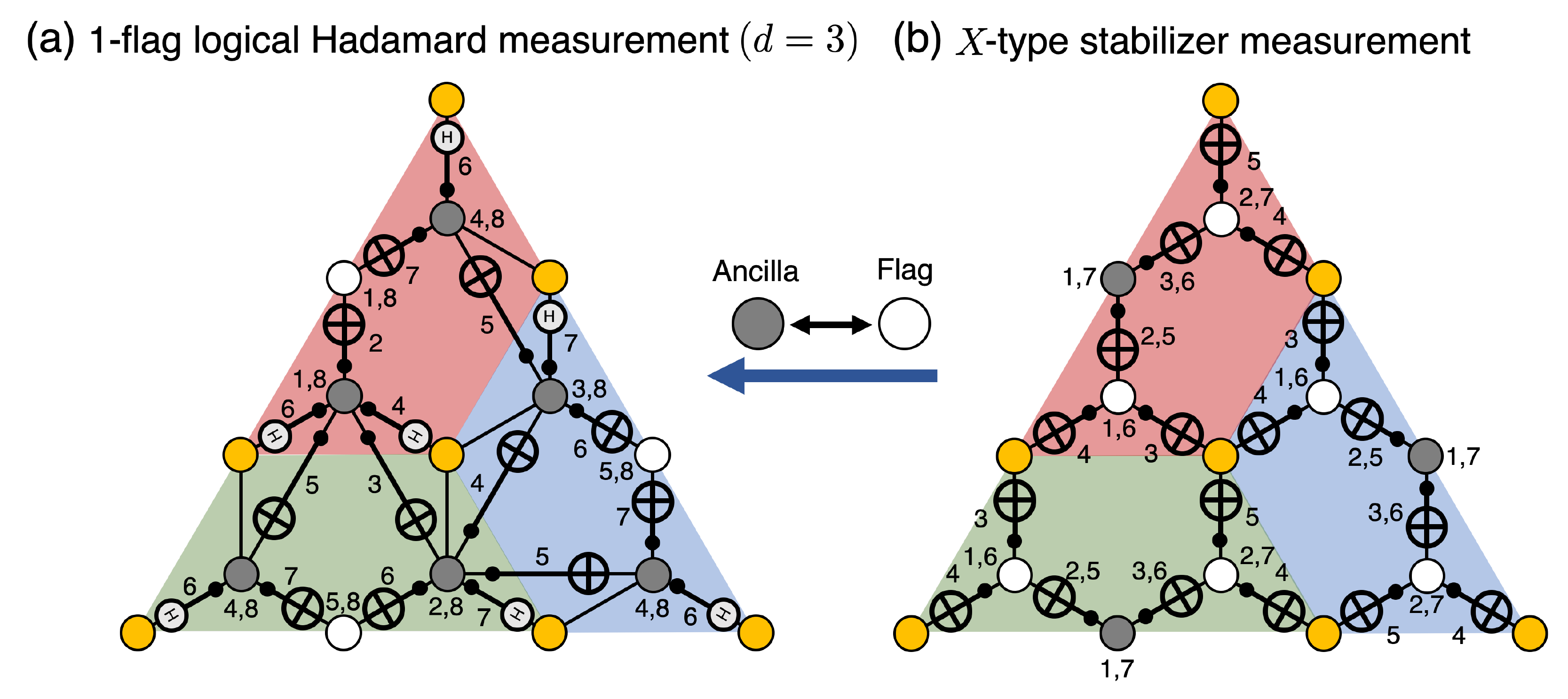}
	\caption{(a) A $1$-flag circuit $H_{m}^{(3)}$ for measuring the logical Hadamard operator $H^{\otimes 7}$ of the $d=3$ triangular color code. (b) A circuit for measuring the X-type stabilizers of the $d=3$ triangular color code. Note that there is a role reversal between the ancilla qubits (grey circles) and the flag qubits (white circles). That is, the ancilla (or flag) qubits in the stabilizer measurement circuit are used as flag (or ancilla) qubits in the Hadamard measurement circuit. Also, the controlled-$H$ symbols in (a) represent the controlled-Hadamard gates. }
	\label{fig:MeasureHd3}
\end{figure*}

Now, one might wonder why three ancilla qubits (grey circles of \cref{fig:ECcircuitFull}b) instead of a single ancilla are used for measuring the weight-six stabilizers (see for instance the weight-six circuits used in Ref.~\cite{CKYZ20} which only require one ancilla). Indeed, if one is only interested in performing fault-tolerant error correction instead of fault-tolerant quantum computation, then a single ancilla qubit suffices since the additional ancillas don't provide more information and they also increase the circuit depth. However in \cref{subsec:TflagHadConstruct}, we show that to use the same qubit layout for measuring the operator $\overline{H} = H^{\otimes n}$ (which is a global operator), the roles of the ancillas and flags in \cref{fig:ECcircuitFull}a are reversed. In other words, the ancilla qubits become flag qubits and the flag qubits become ancilla qubits. To ensure that the circuit $H^{(d)}_{m}$ is a $t = (d-1)/2$-flag circuit for $d \le 7$, we require three flag qubits for every weight-six plaquette and one flag qubit for every weight-four plaquette. Hence if only a single ancilla qubit were used in \cref{fig:ECcircuitFull}b, the circuit $H^{(7)}_{m}$ would not be a three-flag circuit, which would significantly reduce the performance of our protocol (in fact, it would not be possible to achieve the full color code distance for $d>5$). 

To conclude this section, more ancilla qubits than necessary are used to measure the stabilizer generators of the triangular color code. However using the same qubit layout, the extra ancillas (which become flag qubits when measuring $H^{\otimes n}$) ensures that the circuit for measuring the global operator $H^{\otimes n}$ is a $t$-flag circuit (with $t = (d-1)/2$) so that at least $(d+1)/2$ faults are required to produce a logical failure for the scheme in \cref{fig:GeneralHstatePrepScheme}. We refer to this extra redundancy as \textbf{redundant ancilla encoding}. In \cref{subsec:TflagHadConstruct} we provide explicit circuit constructions to measure $H^{\otimes n}$.

\subsection{$t$-flag $H^{(d)}_{m}$ circuit construction for the non-destructive measurement of $\overline{H}$}
\label{subsec:TflagHadConstruct}

\begin{figure*}[t!]
	\centering
	\includegraphics[width=0.85\textwidth]{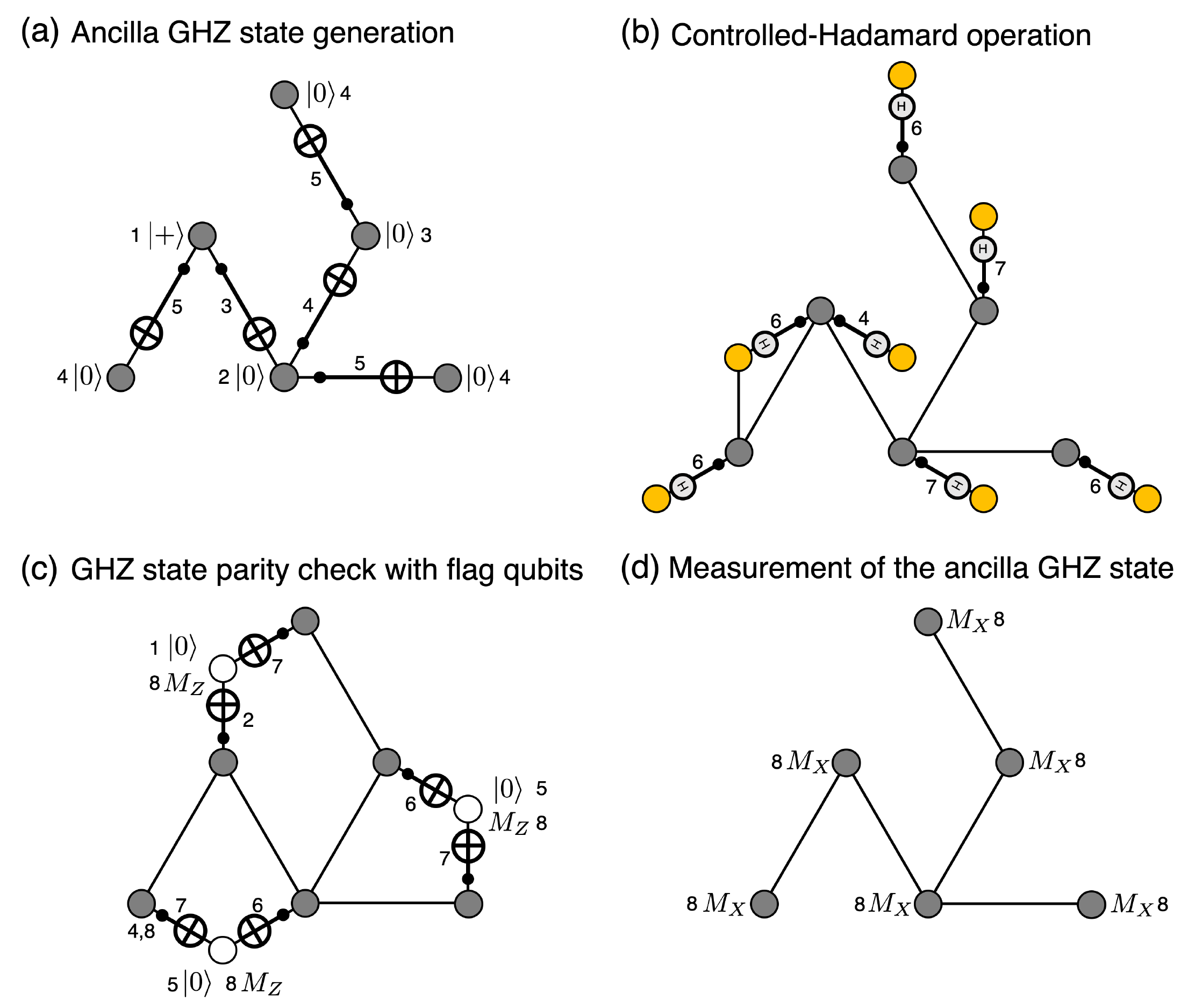}
	\caption{The logical Hadamard circuit ($H^{(3)}_{m}$) in \ref{fig:MeasureHd3}a can be decomposed into four elements, i.e., (a) preparation of the $6$-qubit GHZ state $\frac{1}{\sqrt{2}}( |0\rangle^{\otimes 6} +|1\rangle^{\otimes 6} )$ (or the logical plus state of the $6$-qubit repetition code) in the ancilla qubits (grey circles), (b) controlled-Hadamard gates between the ancilla qubits (in the GHZ state) and the data qubits (yellow circles), (c) parity checks of the ancilla GHZ state by using flag qubits (white circles), and finally (d) measurement of the logical X operator $X^{\otimes 6}$ of the $6$-qubit repetition code on the ancilla qubits.  }
	\label{fig:MeasureHd3Detailed}
\end{figure*}

Here, we describe in detail how to construct a $t$-flag circuit (with $t = (d-1)/2$) for the non-destructive measurement of the logical Hadamard operator (i.e., $H_{m}^{(d)}$). The circuits we construct apply to distance $d$ triangular color codes with $d \in \{3,5,7 \}$. In \cref{fig:MeasureHd3}a, we present a $1$-flag circuit for measuring of the logical Hadamard operator $H^{\otimes 7}$ of the $d=3$ triangular color code. Note that we used the same qubit layout as the one used for the stabilizer measurement circuit ($EC^{(3)}$ circuit shown in \cref{fig:MeasureHd3}b) but there is a role reversal between the ancilla qubits (grey circles) and the flag qubits (white circles). That is, the ancilla qubits in the stabilizer measurement circuit are used as flag qubits in the logical Hadamard measurement circuit and vice versa.  

We now explain why the circuit in \cref{fig:MeasureHd3}a performs a non-destructive measurement of the logical Hadamard operator $H^{\otimes 7}$. Note that the ancilla qubits (grey circles) in \cref{fig:MeasureHd3}a are prepared in the logical plus state of the $6$-qubit repetition code (or the $6$-qubit GHZ state $\frac{1}{\sqrt{2}}( |0\rangle^{\otimes 6} +|1\rangle^{\otimes 6} )$) through the CNOT gates between the ancilla qubits (see \cref{fig:MeasureHd3Detailed}a). Then, as shown in \cref{fig:MeasureHd3Detailed}b, the $7$ controlled-Hadamard gates implement a logical controlled-Hadamard gate between the ancilla $6$-qubit repetition code and the data $d=3$ color code. Eventually, the ancilla qubits (in the $6$-qubit repetition code) are measured in the logical X basis via a $X^{\otimes 6}$ measurement (see \cref{fig:MeasureHd3Detailed}d). Hence, the circuits in Figs.\ \ref{fig:MeasureHd3Detailed}a, b, and d implement the simple non-destructive Hadamard measurement circuit in \cref{fig:HadMeasSimple}, except that the ancilla qubits are now encoded in the $6$-qubit repetition code and the data qubits are encoded in the triangular $d=3$ color code.     

The most important element of the logical Hadamard measurement circuit in \cref{fig:MeasureHd3} is the parity check of the ancilla GHZ state by using flag qubits (see \cref{fig:MeasureHd3Detailed}c). Note that the flag qubits (white circles) non-destructively measure the parity of the ancilla GHZ state $Z_{1}Z_{2}$, $Z_{3}Z_{6}$, and $Z_{4}Z_{5}$, or the three stabilizers of the $6$-qubit repetition code (we labeled the qubits from the top to the bottom and from the left to the right). These stabilizer measurements will be trivial if all the CNOT gates in \cref{fig:MeasureHd3Detailed}a are perfect. However, CNOT gate failures can result in non-trivial flag measurement outcomes. In particular, there are several single CNOT gate failure events that can cause a data qubit error of weight-$2$ or higher. For the Hadamard circuit to be $1$-flag, all these failure events should be caught by flag qubits. Indeed, we verify via a comprehensive numerical search that the logical Hadamard circuit in \cref{fig:MeasureHd3}a is a $1$-flag circuit by confirming that if there is a single fault at any location resulting in a data qubit error $E$ with $\text{min}(\text{wt}(E), \text{wt}(E H^{\otimes 7})) > 1$, at least one flag qubit flags.  

\begin{figure*}[t!]
	\centering
	\includegraphics[width=1.03\textwidth]{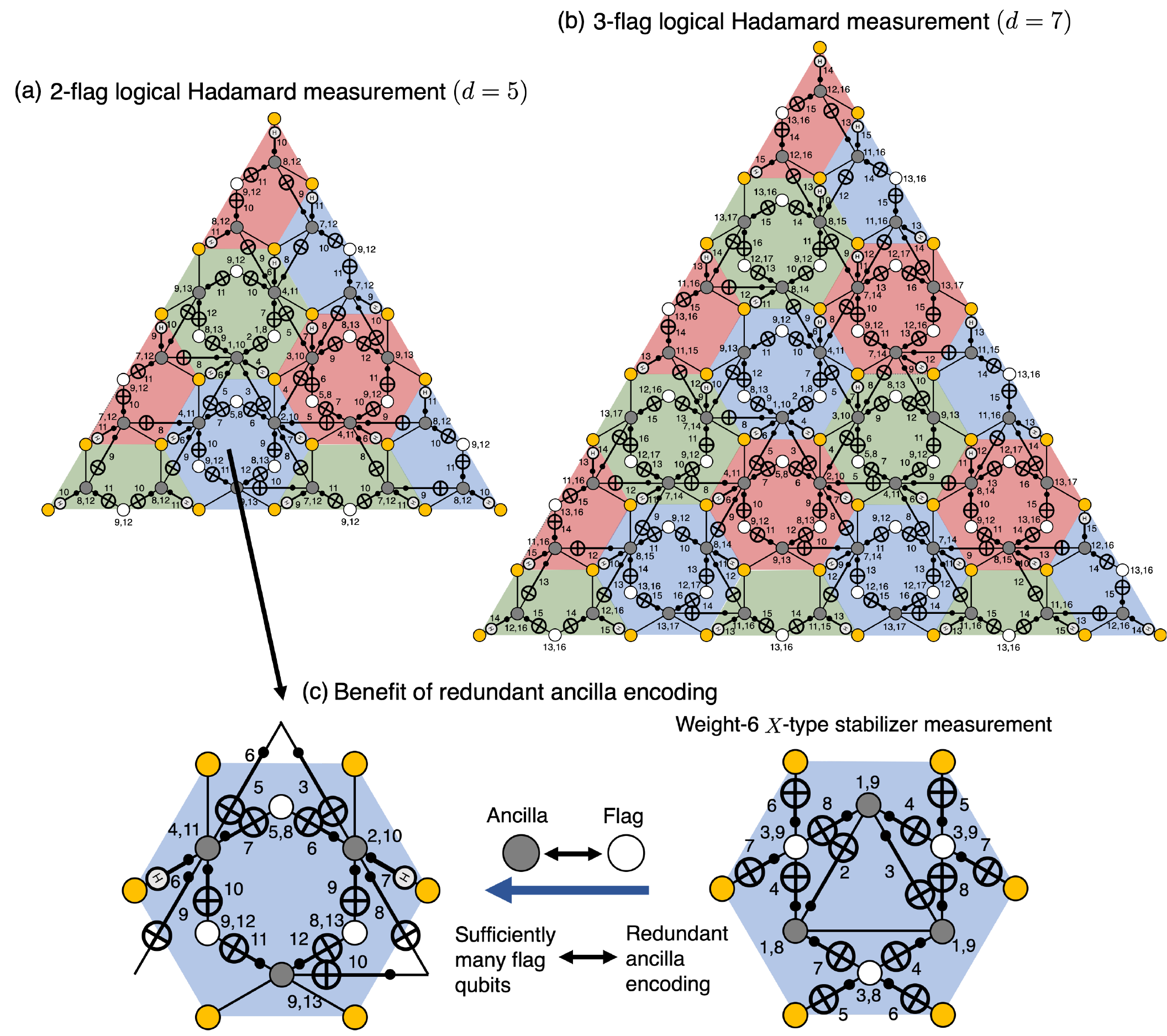}
	\caption{(a) A $2$-flag circuit $H_{m}^{(5)}$ for measuring the logical Hadamard operator $H^{\otimes 19}$ of the $d=5$ triangular color code and (b) a $3$-flag circuit $H_{m}^{(7)}$ for measuring the logical Hadamard operator $H^{\otimes 37}$ of the $d=7$ triangular color code. (c) Illustration of the benefit of using redundant ancilla encoding. Note that the ancilla qubits in the stabilizer measurement circuits are used as flag qubits in the logical Hadamard measurement circuits. Thus, redundant ancilla encoding in the stabilizer measurement circuits allows the logical Hadamard measurement circuits to have sufficiently many flag qubits.  }
	\label{fig:MeasureHd5andd7}
\end{figure*}

Similarly, we construct a $2$-flag circuit for the logical Hadamard measurement $H_{m}^{(d=5)}$ of the $d=5$ triangular color code and a $3$-flag circuit $H_{m}^{(d=7)}$ for the $d=7$ triangular color code (see \cref{fig:MeasureHd5andd7}). The design principle is essentially the same. That is, in the $d=5$ case (i.e., $H_{m}^{(d=5)}$), we first generate a $21$-qubit GHZ state in the ancilla qubits (grey circles), or equivalently, a logical plus state of the $21$-qubit repetition code. The entangled ancilla qubits are then coupled to all of the $19$ data qubits (yellow circles) through controlled-Hadamard gates. Finally, the ancilla qubits (in the $21$-qubit repetition code) are measured in the the logical X basis via a $X^{\otimes 21}$ measurement. Thus, the circuit in \cref{fig:MeasureHd5andd7}a implements a non-destructive measurement of the logical Hadamard measurement of the $d=5$ color code by using ancilla qubits in the $21$-qubit repetition code.  

Precisely because the ancilla qubits are encoded in the $21$-qubit repetition code, we can check if the ancilla qubits are reliably prepared in the logical plus state by measuring $15$ (out of $20$) stabilizers of the ancilla $21$-qubit repetition code using $15$ flag qubits (white circles). Performing an exhaustive numerical search, we can confirm that the circuit $H_{m}^{(d=5)}$ is a $2$-flag circuit. A $3$-flag circuit for the logical Hadamard measurement $H_{m}^{(d=7)}$ of the $d=7$ triangular color code is constructed in the same way by using $45$ ancilla qubits prepared in the logical plus state of the $45$-qubit repetition code, and using $36$ flag qubits checking the $36$ (out of $44$) stabilizers of the $45$-qubit repetition code (see \cref{fig:MeasureHd5andd7}b). We point out that the order and time steps at which the CNOT gates, used to measure the stabilizers of the repetition code with flag qubits, are implemented is very important and must be carefully chosen. 

Recall that we used a redundant ancilla encoding scheme in the stabilizer measurement circuits (see \cref{fig:ECcircuitFull}). Specifically, we redundantly used three ancilla qubits to measure the weight-$6$ stabilizers of the triangular color code. Note that these ancilla qubits are used as flag qubits in the logical Hadamard measurement circuits $H_{m}^{(d)}$ with $d=3,5,7$. Thus, the redundant ancilla encoding in the stabilizer measurement circuits allows the logical Hadamard measurement circuits to have sufficiently many flag qubits while maintaining the same two-dimensional qubit layout. In particular, the logical Hadamard measurement circuits $H_{m}^{(5)}$ and $H_{m}^{(7)}$ have $15$ and $36$ flag qubits that check $15$ (out of $20$) and $36$ (out of $45$) stabilizers of the ancilla GHZ state, respectively. Without the redundant ancilla encoding, we would have had $9$ flag qubits in the $d=5$ case and $18$ flag qubits in the $d=7$ case, as opposed to $15$ and $36$ flag qubits, respectively. We remark that the logical Hadamard measurement circuits $H'^{(5)}_{m}$ and $H'^{(7)}_{m}$ that are constructed with such fewer flag qubits are not $2$-flag and $3$-flag circuits. Thus, the redundant ancilla encoding scheme plays a crucial role in guaranteeing the desired fault-tolerance property of the logical Hadamard measurement circuits.  

\section{Resource overhead for preparing encoded $\ket{\overline{H}}$ states with physical Clifford gates}
\label{sec:Numerics}

\begin{table*} 
	\begin{centering}
		\begin{tabular}{|c|c|c|c|c|c|c|}
			\hline 
			$\ket{\overline{H}}$ (physical Clifford's) & $p$ & $p^{(d)}_{L}$ & $\langle n^{(d)}_{\text{tot}} \rangle$ & $\text{min}(n^{(d)}_{\text{tot}})$ & $s^{(d)}_{O}(p)$ \\
			\hline\hline
			$d=3$ & $10^{-4}$ & $3.45 \times 10^{-6}$ & 17 & 16 & 594   \\
			
			$d=5$ &$10^{-4}$ & $3.6\times 10^{-8}$  & 68 &  55 & 5,694 \\
			
			$d=7$ &$10^{-4}$ & $*4.9\times 10^{-10}$  & 231 & 118 & 27,431 \\	
			\hline \hline
			$d=3$ & $2 \times 10^{-4}$ & $1.39 \times 10^{-5}$ & 17 & 16  & 611  \\
			
			$d=5$ &$2 \times 10^{-4}$ & $3.01\times 10^{-7}$  & 84 & 55 & 7,010 \\
			
			$d=7$ &$2 \times 10^{-4}$ & $*7.83\times 10^{-9}$  & 449 & 118 & 53,359 \\
			\hline \hline
			$d=3$ & $3 \times 10^{-4}$ & $3.11 \times 10^{-5}$ & 18 & 16 & 630  \\
			
			$d=5$ &$3 \times 10^{-4}$ & $1.10\times 10^{-6}$  & 103  & 55 & 8,648\\
			
			$d=7$ &$3 \times 10^{-4}$ & $*3.97\times 10^{-8}$  & 870 & 118 & 103,500 \\
			 \hline \hline
			$d=3$ & $4 \times 10^{-4}$ & $5.64 \times 10^{-5}$ & 18  & 16 & 650 \\
			
			$d=5$ &$4 \times 10^{-4}$ & $2.48\times 10^{-6}$  & 127 & 55 & 10,656 \\
			
			$d=7$ &$4 \times 10^{-4}$ & $*1.25\times 10^{-7}$  & 1,700 & 118 & 202,268\\
			\hline \hline
			$d=3$ & $5 \times 10^{-4}$ & $8.51 \times 10^{-5}$ & 19 & 16 & 670  \\
			
			$d=5$ &$5 \times 10^{-4}$ & $5.23\times 10^{-6}$  & 156 & 55 & 13,115 \\
			
			$d=7$ &$5 \times 10^{-4}$ & $*3.06\times 10^{-7}$  & 3,312 & 118 & 394,177 \\
			 \hline
		\end{tabular}
		\par\end{centering}		
	\caption{\label{tab:PhysicalDirectStatePrep} Logical error rate $p_{L}$, average number of qubits $\langle n^{(d)}_{\text{tot}} \rangle$ (see \cref{eq:LogFailRelation,eq:AverageNumFinal}), minimum number os qubits (\cref{eq:MinNumFinal}) and the space-time overhead (\cref{eq:SpaceTimePhysical}) of the $\ket{\overline{H}}$ state preparation scheme of \cref{sec:FaultTolHstate} obtained from $10^{9}$ Monte-Carlo simulations using the noise model of \cref{subsec:TrainColor} and simulation methods described in \cref{sec:Numerics}. For $p=10^{-4}$, only 68 and 231 qubits are required to prepare $\ket{\overline{H}}$ states with $p_{L} = 3.6\times 10^{-8}$ and $p_{L} = 4.9\times 10^{-10}$ respectively. *For $d=7$, we obtained five data points in the interval $p \in [3 \times 10^{-4}, 4\times 10^{-4}]$ and extrapolated the best fit curve using \cref{eq:LogFailRelation} to obtain all the data in this table.}
\end{table*}

In this section we provide the logical failure rates of the states $\ket{\overline{H}}$ prepared using the scheme presented in \cref{sec:FaultTolHstate} and the noise model described in \cref{subsec:TrainColor}. We also provide the average number of qubits required to produce such states. 

Since the entire sequence of operations in \cref{fig:GeneralHstatePrepScheme} fault-tolerantly prepares $\ket{\overline{H}}$ for $d \le 7$, for a physical error rate $p$ (see the description of the circuit-level noise model used in \cref{subsec:TrainColor}), the output state is afflicted by a logical fault with probability
\begin{align}
p^{(d)}_{L} = \alpha p^{(d+1)/2} + \mathcal{O}(p^{(d+3)/2}),
\label{eq:LogFailRelation}
\end{align}
where $\alpha$ counts all the combinations of $(d-1)/2$ faults which lead to acceptance of our scheme while resulting in a logical $\overline{X}$, $\overline{Y}$ or $\overline{Z}$ error. Hence 
\begin{align}
\alpha < \sum^{N}_{k = 0} \binom{N}{\frac{d-1}{2} + k},
\label{eq:Alpha}
\end{align}
where $N$ is the total number of locations which can fail in the combined circuits of \cref{fig:GeneralHstatePrepScheme}. Note that \cref{eq:Alpha} is a strict inequality since there are a lot of benign locations in the circuits used to prepare $\ket{\overline{H}}$. For values of $p \le 10^{-3}$, the higher order terms in \cref{eq:LogFailRelation} were found to have a negligible impact on $p_{L}$.

\begin{figure}
	\centering
	\includegraphics[width=0.4\textwidth]{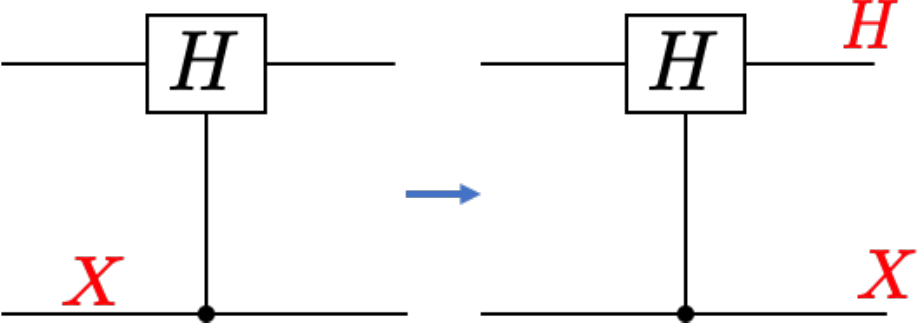}
	\caption{An $X$ error on the control qubit of the controlled-Hadamard gate results in an $H$ error on the target qubit. }
	\label{fig:HasErrProp}
\end{figure}

\begin{figure*}
	\centering
	\includegraphics[width=1.0\textwidth]{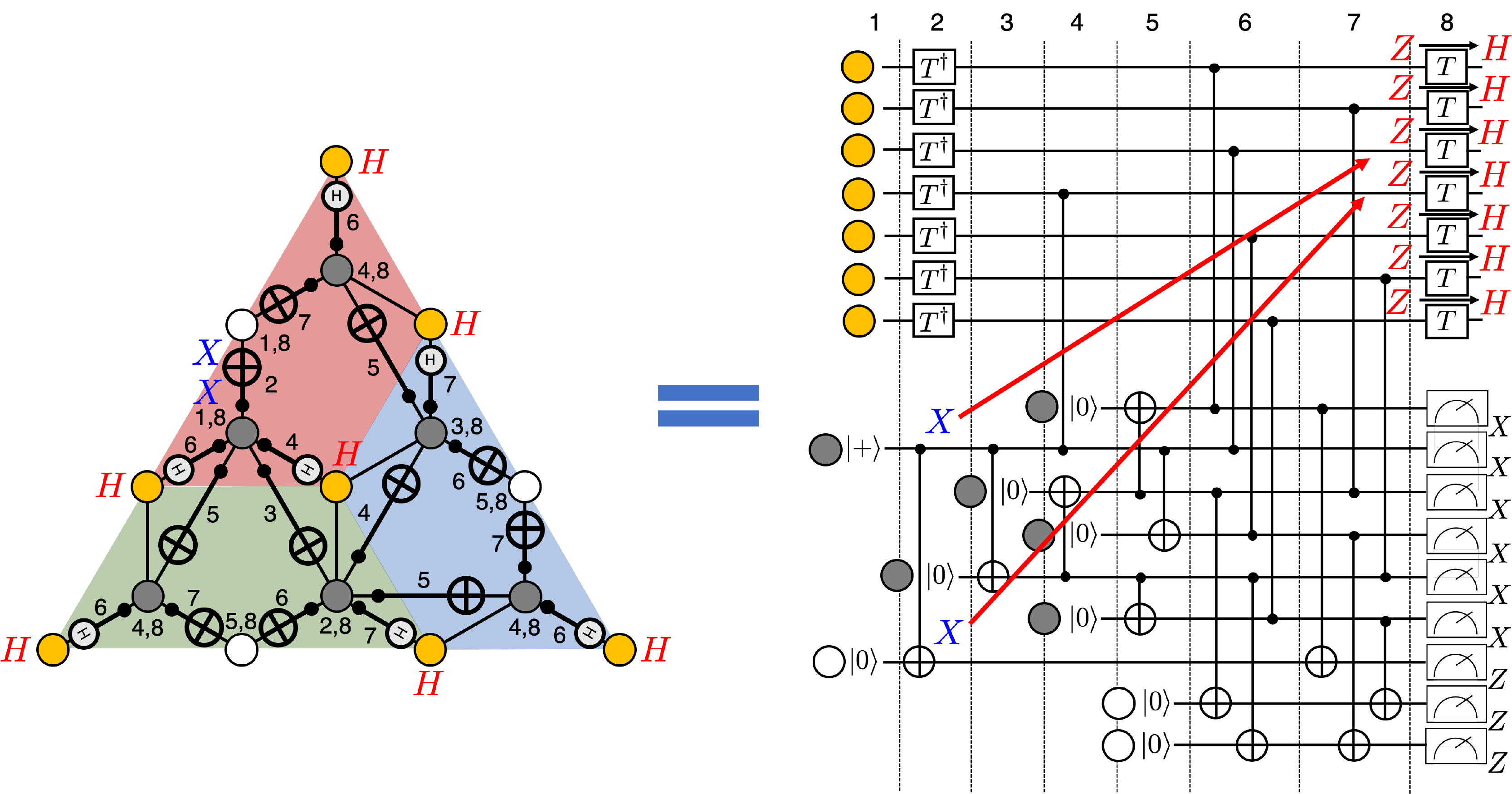}
	\caption{Example of an $X \otimes X$ error on a CNOT gate of the circuit $H^{(3)}_m$ resulting in a $\overline{H} = H^{\otimes n}$ data qubit error without any flag qubits flagging. However, such an error acts trivially on the $\ket{\overline{H}}$ state being prepared.}
	\label{fig:ExampleBenignEvent}
\end{figure*}

Given the acceptance probability $p^{(d)}_{\text{acc}}(p)$ for preparing a distance $d$ encoded state $\ket{\overline{H}}$ with physical error rate $p$, the average number of qubits is given by
\begin{align}
\langle n^{(d)}_{\text{tot}}(p) \rangle = \frac{n^{(d)} + n^{(d)}_{\text{anc}}}{p^{(d)}_{\text{acc}}(p)},
\end{align}
where $n^{(d)} = (3d^2 + 1)/4$ is the number of data qubits, and $n^{(d)}_{\text{anc}}$ is the total number of ancilla and flag qubits used in the circuits $H^{(d)}_{m}$ and $EC^{(d)}$. Since each weight-six stabilizer generator requires six qubits, and each weight-four stabilizer three qubits, we have
\begin{align}
n^{(d)}_{\text{anc}} = 6n^{(d)}_{w_6} +  3n^{(d)}_{w_4},
\end{align}
where $n^{(d)}_{w_4} = (3/2)(d-1)$ and $n^{(d)}_{w_6} = (3d^2 - 12d + 9)/8$ are the number of weight-four and weight-six stabilizers. Putting everything together, we obtain
\begin{align}
\langle n^{(d)}_{\text{tot}} \rangle = \frac{6d^2- 9d + 5}{2p^{(d)}_{\text{acc}}(p)}.
\label{eq:AverageNumFinal}
\end{align}
Note that if we repeat the protocol for preparing $\ket{\overline{H}}$ until the output state is accepted (at the expense of having a higher time cost), the number of qubits required to prepare $\ket{\overline{H}}$ is simply
\begin{align}
\text{min}(n^{(d)}_{\text{tot}}) = \frac{6d^2- 9d + 5}{2}.
\label{eq:MinNumFinal}
\end{align}

We now consider the space-time overhead for implementing the scheme in \cref{sec:FaultTolHstate} where additional qubits (as in \cref{eq:AverageNumFinal}) are used to minimize the time cost. In particular, the space-time overhead $s^{(d)}_{O}(p)$ is given as
\begin{align}
s^{(d)}_{O}(p) = \langle n^{(d)}_{\text{tot}}(p) \rangle(14 + \frac{(d-1)}{2}(t^{(d)}_{H_m} + t^{(d)}_{\text{EC}})),
\label{eq:SpaceTimePhysical}
\end{align}
where $t^{(d)}_{H_m}$ and $t^{(d)}_{\text{EC}}$ are the total number of time steps required to implement one round of the circuits $H^{(d)}_m$ and $EC^{(d)}$ respectively. The factor of 14 in \cref{eq:SpaceTimePhysical} comes from the 14 times steps required to implement the circuit $\ket{\overline{H}}_G$. Note that we pessimistically assume that all time steps are reached when implementing the circuits in \cref{fig:GeneralHstatePrepScheme}. This is pessimistic since when the magic state preparation scheme does not pass the verification steps, fewer than $14 + \frac{(d-1)}{2}(t_{H_m} + t_{\text{EC}})$ time steps are used. 

It is important to point out that due to the presence of $T$ gates (which are non-Clifford) used to implement the controlled-Hadamard gates (see \cref{fig:CHdecomp}), an efficient Monte-Carlo simulation of the circuits $H^{(d)}_{m}$ using Gottesman-Knill error propagation \cite{Gottesman99,AG04} is not possible. Consequently, we divided the Monte-Carlo simulation to calculate $p^{(d)}_{L}$ and $p^{(d)}_{\text{acc}}(p)$ into two parts. First, we perform a Monte-Carlo simulation of the circuit used to prepare $\ket{\overline{H}}_G$. If the output error $E_{\text{out}}$ has a non-trivial syndrome or is a logical operator, the protocol of \cref{sec:FaultTolHstate} is aborted, otherwise, we proceed to simulate the $H^{(d)}_{m}$ and $EC^{(d)}$ circuits. We define $p^{(d)}_{\text{acc,1}}$ to be the probability of proceeding to the $H^{(d)}_{m}$ and $EC^{(d)}$ circuits. Note that there could be other faults in the $H^{(d)}_{m}$ and $EC^{(d)}$ circuits that would cause the protocol to be accepted even though $E_{\text{out}}$ had a non-trivial syndrome or was a logical operator. However such an event would require at least $(d-1)/2$ faults, and since the large majority of malignant error locations are found in $H^{(d)}_{m}$ and $EC^{(d)}$, such an approximation only affects $\alpha$ in \cref{eq:LogFailRelation} by a small constant factor. 

Now, to simulate the circuit $H^{(d)}_{m}$, we use the fact that $T^{\dagger}XT = \frac{1}{\sqrt{2}}(X+Z) = H$ and $T^{\dagger}ZT = \frac{1}{\sqrt{2}}(Z-X) = iYH$. Hence if an $X$ or $Z$ error is input to a $T$ or $T^{\dagger}$ gate, we pessimistically apply an $X$ or $Z$ error to the output, each with $50 \%$ probability. Such an approximation would be exact if twirling operations were performed both before and after the $T^{\dagger}$ and $T$ gates (see \cref{appendix:TwirlingApprox}). To be clear, we do not propose applying twirling operations when implementing our scheme as this could reduce the performance. The approximations stated here are performed to allow us to simulate our scheme on a classical computer. 

We also note that an $X$ error propagating through the control qubit of a controlled-Hadamard gate results in a Hadamard error applied to the data (see \cref{fig:HasErrProp}). Therefore, an $X \otimes X$ error on the first CNOT between the ancillas $\ket{+}$ and $\ket{0}$ in the circuit implementing $H^{(d)}_{m}$ results in the error $H^{\otimes n}$ (which acts trivially on $\ket{\overline{H}}$) without any flag qubits flagging (see \cref{fig:ExampleBenignEvent} for an illustration). However, since the controlled-Hadamard is decomposed as in \cref{fig:CHdecomp}, an $X$ error on the control qubit of the controlled-$Z$ gate results in a $Z$ error on the data. Therefore, if we propagate $Z^{\otimes n}$ (arising from the $X \otimes X$ error at the CNOT mentioned above) through the $T$ gates as described above, the output will not be a benign error. As such, prior to the application of the $T$ gates, let $E_{Z}$ be the $Z$ component of the data qubit errors. For instance, if the data qubit errors are $E = Z \otimes X \otimes Y$, then $E_Z = Z\otimes I \otimes Z$. The $Z$ component which is propagated through the $T$ gates is chosen to be $E'_{Z} = \text{min}(\text{wt}(E_{Z}),\text{wt}(E_{Z}Z^{\otimes n}))$. This prevents a single fault from causing a logical error without any flag qubits flagging in our simulations. 

If all flag qubit and ancilla qubit measurement outcomes in the $(d-1)/2$ applications of the $H^{(d)}_{m}$ and $EC^{(d)}$ circuits are trivial, the output state is accepted. We define $p^{(d)}_{\text{acc,2}}$ to be the probability of acceptance for the second part of the simulation (i.e. the simulation of the $H^{(d)}_{m}$ and $EC^{(d)}$ circuits) . Hence the total acceptance probability is $p^{(d)}_{\text{acc}} = p^{(d)}_{\text{acc,1}}p^{(d)}_{\text{acc,2}}$. To determine if the output error of an accepted state is correctable, we perform one round of perfect error correction using the the \texttt{Lift} decoder.  

\begin{figure*}[t!]
\centering
\includegraphics[width=1.07\textwidth]{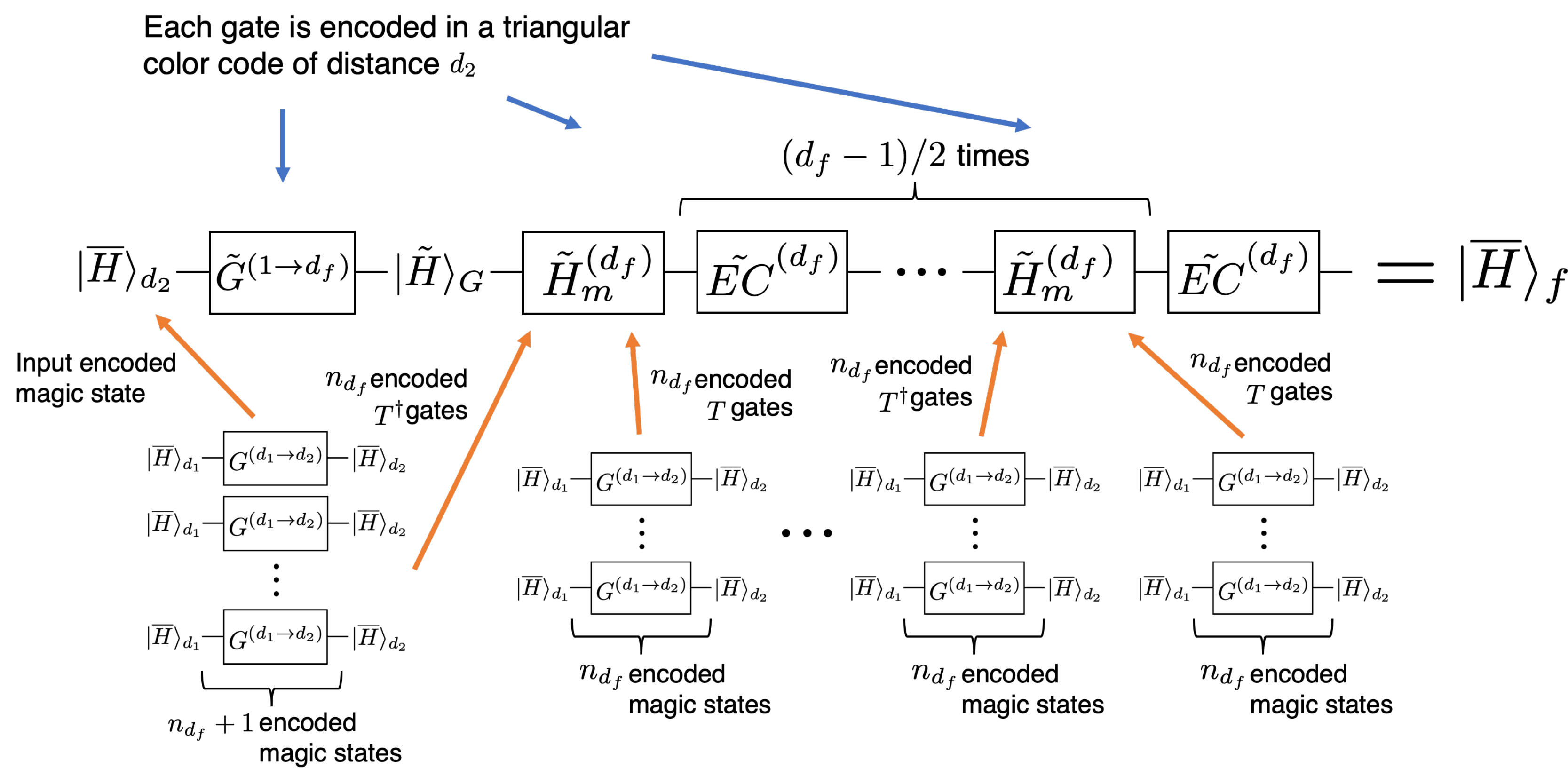}
\caption{General scheme for fault-tolerantly preparing an encoded magic state. Unlike in \cref{fig:GeneralHstatePrepScheme}, each gate is encoded in a triangular color code of distance $d_{2}$. Magic states that are encoded in a triangular color code of distance $d_{1}$ (i.e., $|\overline{H}\rangle_{d_{1}}$) are directly prepared by using the scheme in \cref{fig:GeneralHstatePrepScheme}. These magic states are then grown to the distance $d_{2}$ triangular color code (i.e., $|\overline{H}\rangle_{d_{2}}$) by using the growing scheme in \cref{fig:GeneralGrowingScheme}. Initially, $m_{d_f} \ge n_{d_{f}}+1$ encoded magic states $|\overline{H}\rangle_{d_{2}}$ are prepared where $n_{d_{f}} \equiv (3d_{f}^{2}+1)/4$ (see \cref{eq:mdfProb} for the definition of $m_{d_f}$). One of these encoded magic states is further grown to $|\tilde{H}\rangle$ via a growing circuit $\tilde{G}^{1\rightarrow d_{f}}$ where each gate is encoded in the distance $d_{2}$ color code. The remaining $n_{d_{f}}$ encoded magic states are used to implement the $n_{d_{f}}$ encoded $T^{\dagger}$ gates in the $\tilde{H}_{m}^{(d_{f})}$ circuit. While the circuit $\tilde{H}_{m}^{(d_{f})}$ is being implemented, another $n_{d_{f}}$ encoded magic states need to be prepared so that they can be used to implement $n_{d_{f}}$ encoded $T$ gates at the end of the $\tilde{H}_{m}^{(d_{f})}$ circuit. Since the circuit $\tilde{H}_{m}^{(d_{f})}$ is repeated $(d_{f}-1)/2$ times, we need in total $ (d_{f}-1)n_{d_{f}} +1$ encoded magic states $|\overline{H}\rangle_{d_{2}}$. The $\tilde{EC}^{(d_f)}$ circuits are used to measure the stabilizers of the triangular color codes but with encoded $d_2$ stabilizer operations.}
\label{fig:HadStatePrepFullEncodedClifford}
\end{figure*}

The values of $p^{(d)}_{L}$, $\langle n^{(d)}_{\text{tot}} \rangle$, $\text{min}(n^{(d)}_{\text{tot}})$ and $S^{(d)}_{O}(p)$ for $p \in [10^{-4}, 5\times 10^{-4}]$ are given in \cref{tab:PhysicalDirectStatePrep} and were obtained by performing $10^{9}$ Monte-Carlo simulations on AWS clusters. For values of $p$ resulting in very low logical failure rates, extrapolation of the best fit curve (using \cref{eq:LogFailRelation}) was used to compute the logical error rate of $\ket{\overline{H}}$. Note that using the $d=5$ version of the protocol in \cref{sec:FaultTolHstate}, an $\ket{\overline{H}}$ state can be prepared with logical failure rate $3.6 \times 10^{-8}$ with using only 68 qubits on average when $p = 10^{-4}$. Alternatively, one may use $\text{min}(n^{(d=5)}_{\text{tot}}) = 55$ qubits and repeat the protocol until it is accepted (with an acceptance probability $p^{(d=5)}_{\text{acc}}(p=10^{-4}) = 0.81$). Further, one can use the $d=7$ version of the protocol to produce an $\ket{\overline{H}}$ state with logical failure rate $4.9 \times 10^{-10}$ with only 231 qubits on average. Similarly as above, one may use $\text{min}(n^{(d=7)}_{\text{tot}}) = 118$ qubits and repeat the protocol until it is accepted (with an acceptance probability $p^{(d=7)}_{\text{acc}}(p=10^{-4}) = 0.51$).  In comparison, when $p  =10^{-4}$, the non fault-tolerant magic state distillation scheme in Ref.~\cite{Litinski19magicstate} requires 810 and 1150 qubits to prepare magic states with failure rates $4.4 \times 10^{-8}$ and $9.3 \times 10^{-10}$, respectively. We note that recent works suggests that $p=10^{-4}$ is the appropriate regime to have small enough decoding hardware requirements \cite{DelfosseLowp20}.

An important remark is that the space-time overhead values obtained in Ref.~\cite{Litinski19magicstate} cannot directly be compared with those of \cref{tab:PhysicalDirectStatePrep}. Roughly speaking, one would need to multiply the numbers obtained in  Ref.~\cite{Litinski19magicstate} by at least $6$ (where the factor of 6 comes from the fact that 6 time steps are required to measure all the surface code stabilizers).  To be clear, \cref{eq:SpaceTimePhysical} takes into account the total number of time steps required for each measurement cycle. As such, our scheme provides a space-time overhead improvement, say to obtain a magic state with $p^{(5)}_L = 3.6 \times 10^{-8}$, by at least an order of magnitude. One of the main reasons for the reduction in overhead is that we did not need to use encoded Clifford operations due to the fault-tolerant properties of our circuits. In Ref.~\cite{Litinski19magicstate}, the magic state being distilled needed to be encoded in the $d=7$ surface code (so that the logical error rates of the encoded Clifford gates are approximately $10^{-9}$) to obtain an output state with failure probability $4.4 \times 10^{-8}$. On the other hand, our scheme does not require encoded Clifford gates with very low error rates, and instead works with physical Clifford gates with an error rate $p = 10^{-4}$.
 
 \begin{figure*}
	\centering
	(a)\includegraphics[width=0.45\textwidth]{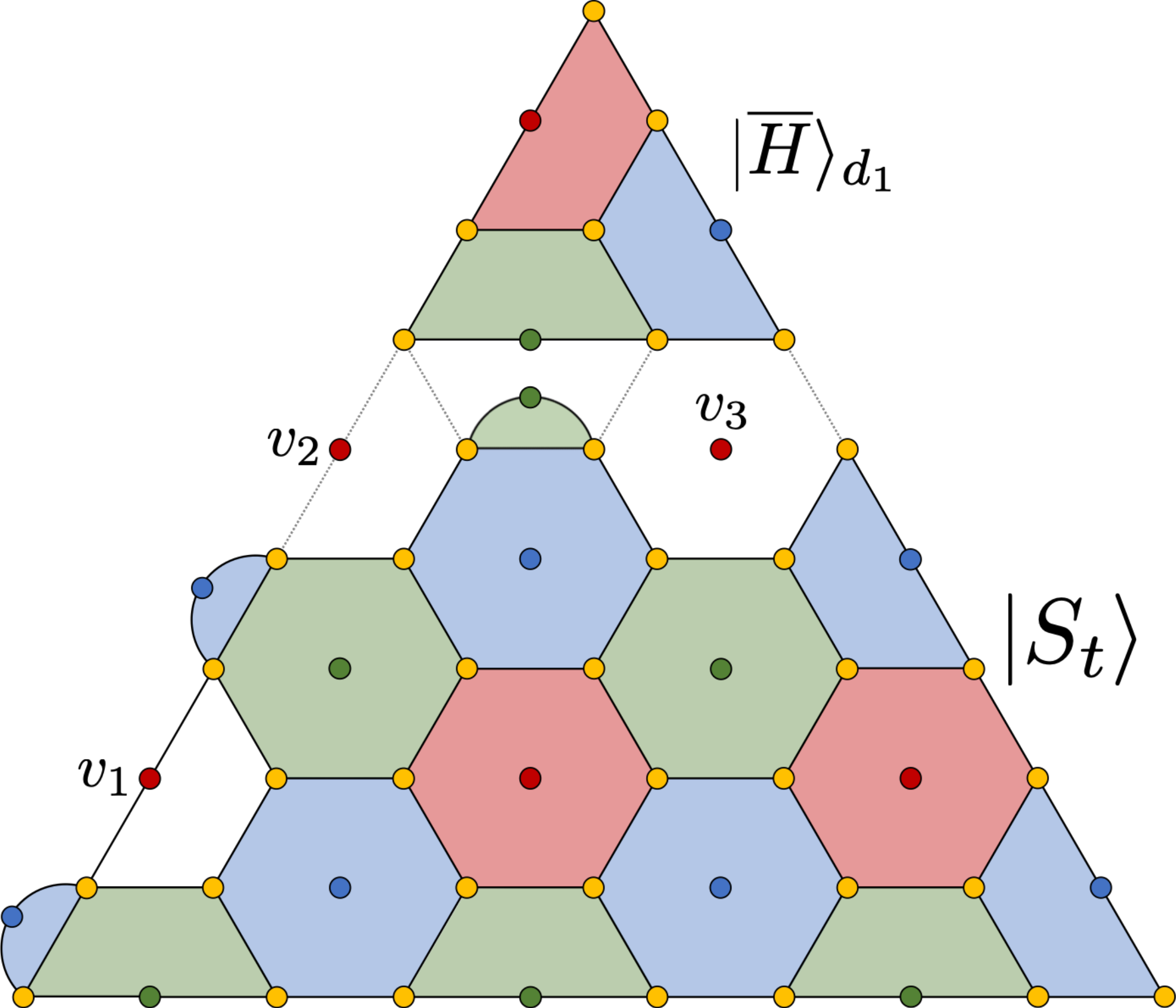}\hspace*{15mm}
         (b)\includegraphics[width=0.25\textwidth]{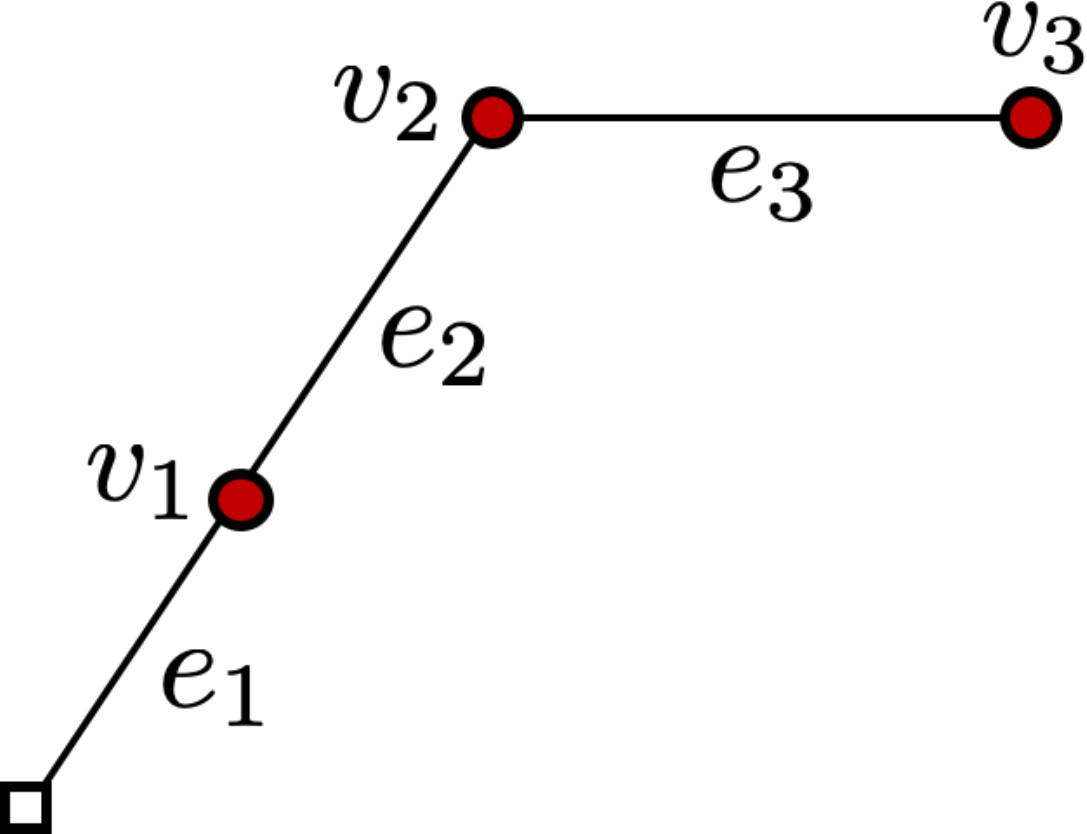}
\caption{(a) Circuit $G^{(3 \rightarrow 7)}$ for growing an $\ket{\overline{H}}$ state encoded in a $d = 3$ triangular color code to one encoded in a $d=7$ triangular color code. A stabilizer state $\ket{S_t}$ is first prepared, and the $d=3$ $\ket{\overline{H}}$ state is prepared following the methods of \cref{sec:FaultTolHstate}. Note that the weight-six checks of $\ket{S_t}$ are implemented using the circuit in \cref{fig:WeightSixCheckOld}. One then measures the $X$ and $Z$-type operators supported on the white plaquettes (also using the circuits in \cref{fig:WeightSixCheckOld} for the weight-six checks) which anti-commute with the weight-two generators of $\mathcal{S}_{\text{st}}$ resulting in random measurement outcomes. The measurements of all plaquettes are repeated three times to distinguish measurement errors from random measurement outcomes and to correct errors. (b) Matching graph ($G^{(7)}_{3x}$ and $G^{(7)}_{3z}$) used to implement the weight-two corrections arising from random measurement outcomes of the operators supported on the white plaquettes. As in \cref{fig:GrowingPhysical}, each edge corresponds to two qubits supported on a weight-two generator of $\mathcal{S}_{\text{st}}$. }
	\label{fig:GeneralGrowingScheme}
\end{figure*}
 
\section{Preparing $\ket{\overline{H}}$ with encoded stabilizer operations}
\label{sec:LogicCliffordStatePrep}

In \cref{sec:Numerics} we showed that using physical Clifford operations which fail with probability $p = 10^{-4}$, $\ket{\overline{H}}$ states with logical error rates near $10^{-8}$ and $10^{-10}$ can be prepared with 68 and 231 qubits on average. However, for many quantum algorithms,  $\ket{\overline{H}}$ states with even lower logical error rates are required. This can be accomplished by encoding all stabilizer operations in a triangular color code since such encoded operations have much lower error rates compared to physical unencoded stabilizer operations. We now describe the implementation of our $\ket{\overline{H}}$ preparation scheme presented in \cref{sec:FaultTolHstate} with encoded stabilizer operations. In what follows, we denote $\ket{\overline{H}}_f$ as the state produced when implementing the scheme of \cref{sec:FaultTolHstate} with encoded stabilizer operations. Further, the circuits $G^{(1 \rightarrow d)}$, $H^{(d)}_{m}$ and $EC^{(d)}$ will be denoted as $\tilde{G}^{(1 \rightarrow d)}$, $\tilde{H}^{(d)}_{m}$ and $\tilde{EC}^{(d)}$ when implemented with encoded stabilizer operations (see \cref{fig:HadStatePrepFullEncodedClifford}). 

Suppose that in order to prepare an $\ket{\overline{H}}_f$ state with some target logical error rate, we require all stabilizer operations to be encoded in a triangular color code of distance $d_2$. First, $\ket{\overline{H}}_{d_1}$ states are prepared using the scheme described in \cref{sec:FaultTolHstate} (with $d_1 \in \{ 3,5,7 \}$ and with physical stabilizer operations). The distance $d_1$ is chosen such that the $\ket{\overline{H}}_{d_1}$ states have smaller logical failure rates compared with those of the distance $d_2$ encoded stabilizer operations. The $\ket{\overline{H}}_{d_1}$ states are used to implement the logical $T$ gates (see \cref{fig:TgateCirc}) and for injection in the circuit  $\tilde{G}^{(1 \rightarrow d_f)}$.  If $d_1 < d_2$, the $\ket{\overline{H}}_{d_1}$ states must first be grown into $\ket{\overline{H}}_{d_2}$ states using a technique analogous to the one illustrated in \cref{fig:GeneralGrowingScheme}. The circuits performing such growing operations are denoted $G^{(d_1 \rightarrow d_2)}$ since all operations are implemented with physical gates. Using encoded $\ket{\overline{H}}_{d_2}$ states ensures that stabilizer operations encoded in a distance $d_2$ triangular color code can be used with the prepared magic states. 

The circuits $G^{(d_1 \rightarrow d_2)}$ are implemented as follows. As in \cref{subsec:GrowingPart}, a stabilizer state $\ket{S_t}$ is prepared, and operators supported on the white plaquettes seperating $\ket{\overline{H}}_{d_1}$ and $\ket{S_t}$ are measured (this step can be viewed as gauge fixing of an underlying subsystem code \cite{VLCABT19}). Measurements of all operators supported on each plaquette of the distance $d_2$ triangular color code are repeated $d_1$ times to correct errors and to distinguish measurement errors from the random outcomes obtained when measuring the white plaquettes. An example for implementing $G^{(3 \rightarrow 7)}$ is provided in \cref{fig:GeneralGrowingScheme}.

Once enough $\ket{\overline{H}}_{d_2}$ states have been prepared (see below), such states are injected into the circuits of \cref{fig:TgateCirc} to perform the logical $T$ gates, in addition to being injected in the circuit $\tilde{G}^{(1 \rightarrow d)}$ (see for instance the circuit used in \cref{fig:GrowingPhysical}, but with distance $d_2$ encoded stabilizer operations). The final $\ket{\overline{H}}_{f}$ state used for computation is then prepared repeating the same steps as in \cref{sec:FaultTolHstate} (i.e. applying $(d_f-1)/2$ pairs of the $\tilde{H}^{(d)}_m$ and $\tilde{EC}^{(d)}$ circuits) with each stabilizer operation encoded in the distance $d_2$ triangular color code. In \cref{fig:HadStatePrepFullEncodedClifford} we provide a schematic illustration of the full scheme described above.

\begin{figure}
	\centering
	\includegraphics[width=0.36\textwidth]{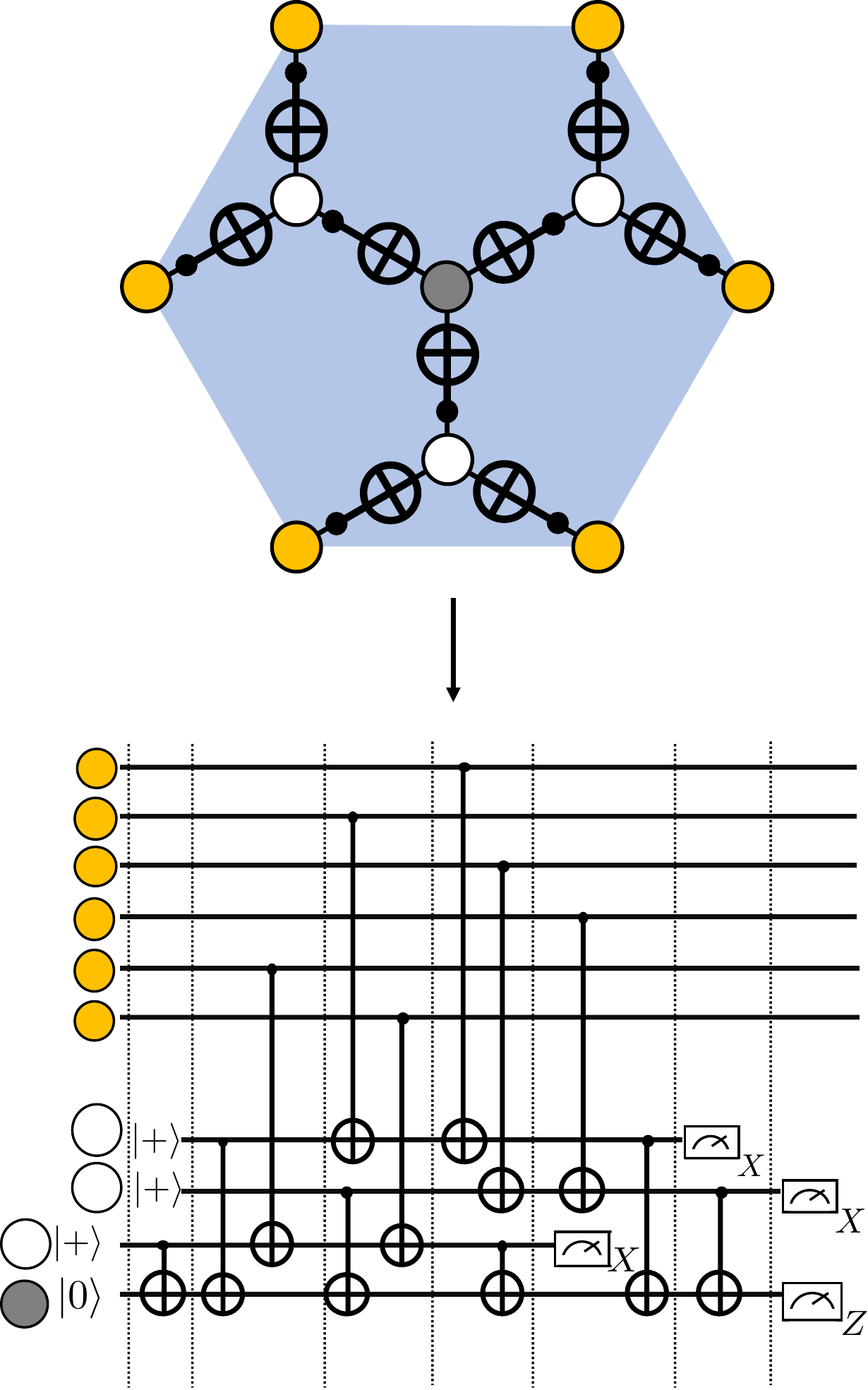}
	\caption{Weight-six $Z$-type stabilizer plaquette used in the lattice $\mathcal{L}$ for performing the encoded Clifford operations. The plaquettes used for preparing the stabilizer state $\ket{S_t}$ which is gauge fixed with $\ket{\overline{H}}_{d_1}$ also use the layout shown in this figure. The $X$-type stabilizer is obtained by inverting the directions of the CNOT gates, swapping $\ket{0}$ with $\ket{+}$ and exchanging the $Z$-basis measurements with $X$-basis measurements.}
	\label{fig:WeightSixCheckOld}
\end{figure}

To compute the overhead for preparing the state $\ket{\overline{H}}_{f}$, we consider the case where all the $T^{\dagger}$ gates are simultaneously implemented during the second time step of the $H^{(d)}_{m}$ circuit. We can thus prepare $m_{d_f}$ $\ket{\overline{H}}_{d_1}$ states which are used for implementing the $T^{\dagger}$ gates in addition to injecting one of these states into the circuit $\tilde{G}^{(1 \rightarrow d)}$. The probability that at least $(3d^2_f+1)/4 + 1 = n_{d_f} + 1$ $\ket{\overline{H}}_{d_1}$ states pass the verification test is given by
\begin{widetext}
\begin{equation}
P^{(n_{d_f})}_{A,d_f}(p,m_{d_f}) = 
\sum_{k = n_{d_f} + 1}^{m_{d_f}} \binom{m_{d_f}}{k} p^{(d_1)}_{\text{acc}}(p)(1 - p^{(d_1)}_{\text{acc}}(p))^{m_{d_f} - k},
\label{eq:mdfProb}
\end{equation}
\end{widetext}
where $p^{(d_1)}_{\text{acc}}(p)$ is the probability of acceptance for preparing the state $\ket{\overline{H}}_{d_1}$. An accepted $\ket{\overline{H}}_{d_1}$ state then grows into an encoded $\ket{\overline{H}}_{d_2}$ state since the Clifford operations are chosen to be encoded in the distance $d_2$ triangular color code. Since the weight-six stabilizers of the stabilizer state (and all encoded Clifford gates) are obtained from the circuit in \cref{fig:WeightSixCheckOld}, the total number of qubits required for the stabilizer state $\ket{S_t}$ is
\begin{align}
n_{S_t}(d_1,d_2) = \frac{(3d_2-1)^2}{4} - \frac{(3d_1-1)^2}{4},
\end{align}
and the number of qubits for each $\ket{\overline{H}}_{d_1}$ state is  
\begin{align}
n_{d_1} = \frac{6d{_1}^2- 9d_1 + 5}{2}.
\end{align}
Lastly, since the qubits in the implementation for preparing $\ket{\overline{H}}_f$ using the protocol of \cref{sec:FaultTolHstate} are encoded in the triangular color code with distance $d_2$, we require an additional
\begin{align}
n_{\text{add}}(d_1,d_f) = \frac{(3d_2-1)^2(6d^2_f-9d_f+5)}{8},
\end{align}
qubits. Hence, the total average number of qubits $\langle n_f \rangle$ required to prepare $\ket{\overline{H}}_f$ is
\begin{widetext}
\begin{equation}
\langle n_f(p,m_{d_f}) \rangle = 
\frac{n_{\text{add}}(d_1,d_f)+m_{d_f}(n_{d_1} + n_{S_t}(d_1,d_2))}{P^{(n_{d_f}+1)}_{A,d_f}(p,m_{d_f})\Big (P^{(n_{d_f})}_{A,d_f}(p,m_{d_f})\Big )^{(d_f - 2)} P_{A,H_f}(p)},
\label{eq:totAve}
\end{equation}
\end{widetext}
where $P_{A,H_f}(p)$ is the acceptance probability for preparing $\ket{\overline{H}}_f$ with Clifford operations encoded in a distance $d_2$ triangular color code. For a fixed value of $p$, $m_{d_f}$ is chosen to minimize \cref{eq:totAve}. Note that we assume that all the qubits used to prepare the $m_{d_f}$ $\ket{\overline{H}}_{d_1}$ states can be reused to implement the $T$ gates at the end of the $H^{(d)}_{m}$ circuit. In doing so, it is assumed that the time scale required to prepare the $\ket{\overline{H}}_{d_2}$ states is less than or equal to the time scale required to implemented all the encoded operations prior to applying the $T$ gates at the end of the $\tilde{H}^{(d)}_{m}$ circuit. Also, the denominator of \cref{eq:totAve} has the factor $P^{(n_{d_f}+1)}_{A,d_f}(p,m_{d_f})\Big (P^{(n_{d_f})}_{A,d_f}(p,m_{d_f})\Big )^{(d_f - 2)}$ for the following reasons: In the first time step, an extra magic state is used in the $\tilde{G}^{(1 \rightarrow d_f)}$ circuit (see \cref{fig:HadStatePrepFullEncodedClifford}). Second, only $n_{d_f}$ $\ket{\overline{H}}_{d_2}$ states are required when implementing the sequence of $T$ gates (which are implemented after the $T^{\dagger}$ gates). The term $P^{(n_{d_f})}_{A,d_f}(p,m_{d_f})$ is taken to the power of $d_f - 2$ since the $H^{(d_f)}_m$ circuit is repeated $(d_f - 1)/2$ times (and each circuit requires the injection of $n_{d_f}$ magic states for both the $T^{\dagger}$ and $T$ gates).

If one is willing to significantly increase the time-overhead required for preparing all the magic states used in the preparation scheme of $\ket{\overline{H}}_f$, the number of qubits required to prepare $\ket{\overline{H}}_f$ can be reduced compared to the requirements given by \cref{eq:totAve}. In particular, one can repeat the $\ket{\overline{H}}_{d_1}$ protocol until $n_{d_f}$ magic states are simultaneously accepted and ready to be grown to  $\ket{\overline{H}}_{d_2}$ states. Such qubits are then reused to prepare $\ket{\overline{H}}_{d_1}$ states prior to implementing the $T^{\dagger}$ and $T$ gates every time the $H^{(d)}_m$ circuit is repeated. If any of the $\ket{\overline{H}}_{d_1}$ states do not pass the verification test described in \cref{sec:FaultTolHstate}, the protocol is aborted. In this case, the minimum number of qubits required to prepare $\ket{\overline{H}}_f$ is simply
\begin{align}
\text{min}(n_{f}(p)) = n_{\text{add}}(d_1,d_f)+ (n_{d_f}+1)(n_{d_1} + n_{S_t}(d_1,d_2)).
\label{eq:MinNumQubits}
\end{align}  

We now describe how to compute the space-time overhead for preparing $\ket{\overline{H}}_f$. We first need to consider the space-time overhead of preparing the $m_{d_f}$ $\ket{\overline{H}}_{d_1}$ states, and then growing them to $\ket{\overline{H}}_{d_2}$ states. The space-time overhead for preparing a single $\ket{\overline{H}}_{d_1}$ state is obtain in a similar way to \cref{eq:SpaceTimePhysical} and is given by
\begin{align}
S_1 = n_{d_1}(14 + \frac{(d_1 - 1)}{2}(t^{(d_1)}_{H_m} + t^{(d_1)}_{\text{EC}}),
\end{align}
where $t^{(d_1)}_{H_m}$ and $t^{(d_1)}_{\text{EC}}$ correspond to the number of time steps for implementing the $H^{(d_1)}_{m}$ and $EC^{(d_1)}$ circuits. 
When growing a state $\ket{\overline{H}}_{d_1}$ to $\ket{\overline{H}}_{d_2}$, the measurements of the plaquettes are repeated $d_1$ times, with the maximum number of time steps for each round of measurement being $t^{(d_1)}_{\text{EC}}$ (which come from measuring the stabilizers for the state $\ket{\overline{H}}_{d_1}$ using the circuits of \cref{fig:ECcircuitFull}). Therefore, the space-time overhead for growing an $\ket{\overline{H}}_{d_1}$ to $\ket{\overline{H}}_{d_2}$ state is given by
\begin{align}
S^{(d_2)}_{G} = t^{(d_1)}_{\text{EC}}d_1(n_{S_t}(d_1,d_2) + n_{d_1}).
\end{align}
Since we prepare $m_{d_f}$ such states, the space-time overhead for preparing all of the $\ket{\overline{H}}_{d_2}$ states which are injected in the $T^{\dagger}$ gates and the circuit $\tilde{G}^{(1 \rightarrow d_f)}$ is 
\begin{align}
S_2 = m_{d_f}(S_1 + S^{(d_2)}_{G}).
\label{eq:S2}
\end{align}
The last step consists of implementing the $(d_f-1)/2$ pairs of $\tilde{H}^{(d_f)}_m$ and $\tilde{EC}^{(d_f)}$ circuits. Each implementation of an encoded Clifford gate requires $d_2$ rounds of error correction, and an encoded block consists of $(3d_2 - 1)^2/4$ qubits. Furthermore, if $d_2 > 3$, the total number of time steps for one round of error correction is 16. Thus the space-time overhead for preparing $\ket{\overline{H}}_{f}$ with encoded Clifford gates is
\begin{align}
S_{d_f} &= 16n_{\text{add}}(d_2,d_f)(\text{max}(14,d_2) + \nonumber \\ 
&\frac{(d_f - 1)}{2}(\text{max}(t^{(d_f)}_{H_m},d_2) + \text{max}(t^{(d_f)}_{\text{EC}},d_2)).
\label{eq:Sdf}
\end{align}
Note that in \cref{eq:Sdf}, we choose a large enough time window for the implementation of the circuits $\tilde{G}^{(1 \rightarrow d_f)}$, $\tilde{H}^{(d_f)}_m$ and $\tilde{EC}^{(d_f)}$ to ensure that at least $d_2$ time steps occur allowing one to decode to the full $d_2$ color code distance. Since all logical gates in this time window are Clifford (for $H^{(d)}_m$, this is true for gates applied in between the $T^{\dagger}$ and $T$ gates) and performed using lattice surgery \cite{HFDvM12,LR14,LVO18,FGLatticeSurg18}, one can perform the appropriate Pauli frame updates to incorporate the correct decoding scheme over the full triangular color code cycle. In our numerical simulations, we pessimistically added an error at each logical Clifford gate location using the failure probabilities obtained in Ref.~\cite{CKYZ20} instead of using such probabilities for adding failures over a full distance $d_2$ triangular color code cycle. Hence the $p^{(d_f)}_{L}$ values reported in \cref{tab:LogicalDirectStatePrep} are upper bounds on the exact values that would be obtained using our scheme. 

Combining \cref{eq:S2,eq:Sdf}, the total space-time overhead for the $\ket{\overline{H}}_{f}$ preparation scheme is given by
\begin{align}
S^{(d_f)}_{\text{tot}} = \frac{S_2+S_{d_f}}{P^{(n_{d_f}+1)}_{A,d_f}(p,m_{d_f})\Big (P^{(n_{d_f})}_{A,d_f}(p,m_{d_f})\Big )^{(d_f - 2)} P_{A,H_f}(p)}.
\label{eq:SpaceTime}
\end{align}

\begin{table*} 
	\begin{centering}
		\begin{tabular}{|l|c|c|c|c|c|c|c|}
			\hline 
			$\ket{\overline{H}}_{f}$ & $\ket{\overline{H}}_{d_1}$ & $p$ & Color code distance ($d_2$) & $p^{(d_f)}_{L}$ & $\langle n_f(p,m_{d_f}) \rangle$ & $\text{min}(n_{f}(p))$ & $S^{(d_f)}_{\text{tot}}$ \\
			\hline\hline
			$d_f=3$ & $d_1=7$ & $10^{-4}$ & $d_2 = 11$ & $3.59 \times 10^{-15}$ & 10,917 & 6,288 & $3.91 \times 10^6$  \\
			\hline \hline
						
			$d_f=5$ & $d_1=5$ & $10^{-4}$ & $d_2 = 9$ & $1.10 \times 10^{-17}$ & 14,955 & 12,795 & $1.22 \times 10^7$ \\	
				
			$d_f=5$ & $d_1=5$ & $2 \times 10^{-4}$ & $d_2 = 11$ & $4.01 \times 10^{-16}$ & 25,265 & 19,320 & $1.88 \times 10^7$ \\
			
			$d_f=5$ & $d_1=5$ & $3 \times 10^{-4}$ & $d_2 = 15$ & $6.12 \times 10^{-17}$ & 53,449 & 36,420 & $3.84 \times 10^7$ \\	 		
			
			\hline \hline
						
			$d_f=7$ & $d_1=3$ & $10^{-4}$ & $d_2 = 7$ & $5.37 \times 10^{-18}$ & 16,262 & 15,600 & $2.30 \times 10^7$ \\	
				
			$d_f=7$ & $d_1=3$ & $2 \times 10^{-4}$ & $d_2 = 9$ & $5.11 \times 10^{-17}$ & 28,141 & 26,364 & $3.91  \times 10^7$ \\
			
			$d_f=7$ & $d_1=3$ & $3 \times 10^{-4}$ & $d_2 = 11$ & $9.87 \times 10^{-17}$ & 43,329 & 39,936 & $5.96  \times 10^7$ \\
			
			$d_f=7$ & $d_1=3$ & $4 \times 10^{-4}$ & $d_2 = 13$ & $5.74 \times 10^{-16}$ & 62,540 & 56,316 & $9.26  \times 10^7$ \\
			
			$d_f=7$ & $d_1=3$ & $5 \times 10^{-4}$ & $d_2 = 15$ & $1.17 \times 10^{-15}$ & 84,200 & 75,504 & $1.156  \times 10^8$ \\
			
			 \hline
		\end{tabular}
		\par\end{centering}		
	\caption{\label{tab:LogicalDirectStatePrep} Qubit and space-time overhead of various schemes using encoded Clifford gates to obtain $\ket{\overline{H}}_f$ states with logical error rates $p^{(d_f)}_L < 4 \times 10^{-15}$. Here $d_2$ is the triangular color code distance used to encode the logical Clifford gates, $d_f$ is distance used for the $\ket{\overline{H}}$ state preparation scheme of \cref{sec:FaultTolHstate}, $d_1$ is the distance of $\ket{\overline{H}}$ prior to being grown into $\ket{\overline{H}}_{d_2}$ and $p$ is the physical error rate (see \cref{subsec:TrainColor}). We provide both the average number of qubits (given by \cref{eq:totAve}) and also the minimum number of qubits (see \cref{eq:MinNumQubits}) for preparing $\ket{\overline{H}}_f$. The space-time overhead is given by \cref{eq:SpaceTime}.}
\end{table*}

\begin{table*} 
	\begin{centering}
		\begin{tabular}{|l|c|c|c|c|c|c|}
			\hline 
			$\ket{\overline{H}}_{f}$ & $\ket{\overline{H}}_{d_1}$ & $p$ & Surface code distance ($d_2$) & $p^{(d_f)}_{L}$ & $\langle n_f(p,m_{d_f}) \rangle$ & $\text{min}(n_{f}(p))$\\
			\hline\hline
			$d_f=7$ & $d_1=3$ & $10^{-4}$ & $d_2 = 5$ & $5.07 \times 10^{-17}$ & 10,025 & 9,506  \\
			$d_f=7$ & $d_1=3$ & $10^{-3}$ & $d_2 = 11$ & $8.11 \times 10^{-20}$ & 60,886 & 47,324   \\
			
			 \hline
		\end{tabular}
		\par\end{centering}		
	\caption{\label{tab:SurfaceCodeLogicStatePrep} Qubit overhead of schemes for obtaining an encoded $\ket{\overline{H}}_f$ as in \cref{tab:LogicalDirectStatePrep}, but with logical stabilizer operations encoded in a distance $d_2$ surface code. Note that the states $\ket{\overline{H}}_{d_2}$ are first encoded in the color code, and lattice surgery is performed to obtain an $\ket{\overline{H}}_{d_2}$ state encoded in the surface code as in \cite{PFB17}.}
\end{table*}

In \cref{tab:LogicalDirectStatePrep} we provide the average number of qubits, minimum number of qubits and space-time overhead required to prepare $\ket{\overline{H}}_{f}$ states with logical failure rates $p^{(d_f)}_{L} < 4 \times 10^{-15}$. To obtain such results, we assume that each Clifford gate encoded in the triangular color code fails according to the logical error rate polynomials obtained in Ref.~\cite{CKYZ20} (which we call $p^{(d_2)}_{LC}(p)$). Hence, when preparing the stabilizer state $\ket{S_t}$ and for all encoded Clifford operations, the ancilla qubit layout for the weight-six checks are chosen as in \cref{fig:WeightSixCheckOld} (note that the weight-four checks remain unchanged). Further, the distance $d_1$ is chosen to be the smallest $d_1$ which ensures that $\ket{\overline{H}}_{d_1}$ has a lower logical error rate than $p^{(d_2)}_{LC}(p)$. Lastly, due to the low failure rates of the encoded components, to obtain $p^{(d_f)}_{L}$ we repeat the simulation described in \cref{sec:Numerics} for physical values of $p > 10^{-3}$, and extrapolate the best fit curves to the regime where $p \in [10^{-4}, 5 \times 10^{-4}]$.

The numerical values obtained in \cref{tab:LogicalDirectStatePrep} shows that the least costly scheme to prepare the state $\ket{\overline{H}}_f$ with $p^{(d_f)}_{L} < 4 \times 10^{-15}$ when $p = 10^{-4}$ is to first prepare the states $\ket{\overline{H}}_{d_1}$ with $d_1 = 7$, and to grow these states to encoded $d_2 = 11$ states. Using distance $d_2 = 11$ encoded stabilizer operations, the final magic state $\ket{\overline{H}}_f$ is prepared using the distance $d_f = 3$ scheme of \cref{sec:FaultTolHstate}. On average, the amount of qubits required to prepare such a state is 10,917 and the space-time overhead is $3.91 \times 10^6$. If the time cost for preparing the $\ket{\overline{H}}_{d_1}$ states is of a lesser concern, the minimum number of qubits required to prepare $\ket{\overline{H}}_f$ with $p^{(d_f)}_{L} < 4 \times 10^{-15}$ is 6,288. To compare with other schemes, in Ref.~\cite{Litinski19magicstate}, a magic state with a logical error failure rate of $2.4 \times 10^{-15}$ required a minimum of 16,400 qubits.

To obtain a state $\ket{\overline{H}}_f$ with $p^{(d_f)}_{L} \approx 10^{-15}$ when $p = 10^{-3}$ using a small amount of resources requires encoded stabilizer operations with much lower logical failure rates than what is achieved with the triangular color code family. One viable option is to use stabilizer operations encoded in the surface code due to the low error rates that can be achieved when $p = 10^{-3}$ \cite{FowlerAutotune}. However, in such a setting, after the states $\ket{\overline{H}}_{d_2}$ have been prepared, they must be teleported to the surface code before they can be injected in the circuit of \cref{fig:TgateCirc} and in the circuit implementing $\tilde{G}^{(1 \rightarrow d_f)}$. In particular, one can convert the color code encoded state to the surface code using lattice surgery techniques as was done in Ref.~\cite{PFB17}. In \cref{tab:SurfaceCodeLogicStatePrep} we provide estimates of the qubit overhead for preparing $\ket{\overline{H}}_f$ when the stabilizer operations are encoded in the surface code. The cost of first encoding the states $\ket{\overline{H}}_{d_2}$ in the color code and then using extra qubits to convert such states into the surface code is taken into account. However, we assume that the quality of the encoded $\ket{\overline{H}}_{d_2}$ states does not change when performing lattice surgery. Although such an omission is optimistic, we verified numerically that when only the $T$ and $T^{\dagger}$ gate locations are allowed to fail, the protocol of \cref{sec:FaultTolHstate} produces $\ket{\overline{H}}_f$ states with logical error rates two to four orders of magnitude (depending on the value of $p$) less than when all stabilizer operations fail according to the noise model described in \cref{subsec:TrainColor}. As such, the simulation provides evidence that the logical error rates obtained in \cref{tab:SurfaceCodeLogicStatePrep} are good estimates of the error rates that would be obtained when considering errors introduced when performing lattice surgery to obtain surface code encoded $\ket{\overline{H}}_{d_2}$ states.

Instead of performing lattice surgery to convert a color code encoded $\ket{\overline{H}}_{d_2}$ state to one encoded in the surface code, another option would be to initially prepare an encoded $\ket{\overline{H}}_{d_2}$ state in a small distance surface code using some other method, such as a magic state distillation protocol. The $\ket{\overline{H}}_{d_2}$ states would then be injected in the $T$ gate circuits of \cref{fig:TgateCirc} in addition to the $\tilde{G}^{(1 \rightarrow d_f)}$ circuit in order to prepare an $\ket{\overline{H}}_{f}$ state using the methods presented in \cref{sec:FaultTolHstate}. To obtain comparable logical failure rates to the ones shown in \cref{tab:LogicalDirectStatePrep} (say $p^{(d_f)}_L \le 10^{-15}$) at $p = 10^{-3}$, the surface code distance $d_2$ would need to be $d_2 = 15$ if the $d_f = 3$ scheme was chosen, $d_2 = 13$ if the $d_f = 5$ scheme was chosen and $d_2 = 11$ if the $d_f = 7$ scheme were chosen. Note that the logical error rate of the surface code is given by $10^{-9}$ for $d_{2}=15$, $10^{-8}$ for $d_{2}=13$, and $10^{-7}$ for $d_{2}=11$ \cite{FowlerAutotune}. In contrast, $d_2 \ge 27$ would be required if conventional non fault-tolerant magic state distillation schemes are used, since in this case the stabilizer operations should have a logical error rate as low as $10^{-15}$. A careful analysis of the overhead would require choosing the appropriate magic state distillation protocol (or some other scheme which uses fault-tolerant circuits to prepare encoded magic states) and therefore such an analysis is left for future work. 

Lastly, one could also directly prepare an encoded $\ket{\overline{H}}_{d_2}$ state in a distance $d_2$ surface code using the non-fault-tolerant state injection methods of Ref.~\cite{LiStateInjection15}. Such states would then be injected in the scheme for preparing $\ket{\overline{H}}_{f}$ (see \cref{fig:HadStatePrepFullEncodedClifford}) with encoded stabilizer operations in a distance $d_2$ surface code. In this case, the injected $\ket{\overline{H}}_{d_2}$ states would have much higher failure rates compared to the encoded stabilizer operations. In such a setting, extending our scheme to $d_f \ge 9$ could potentially be very beneficial. However, to avoid preparing a state $\ket{\overline{H}}_f$ with $d_f > 7$, one could consider a two level approach. As a first step, one could inject $\ket{\overline{H}}_{d_2}$ states to first prepare an $\ket{\overline{H}}_{f}$ state with $d_f \le 5$. Afterwords, the obtained $\ket{\overline{H}}$ could be further injected to prepare a new $\ket{\overline{H}}_{f}$ state with $d_f \ge 5$. 

\section{Conclusion and outlook}
\label{sec:Conclusion}

In this work we showed how to prepare an $\ket{H}$-type magic state in a fault-tolerant way using a two-dimensional color code architecture requiring only nearest-neighbor interactions. The proposed architecture can be used to both measure local stabilizers of the color code in addition to a global operator, without the need for changing the qubit layout. Such an architecture was made possible with the use of flag qubits, in addition to a new concept which we call redundant ancilla encoding. Estimating the performance of our scheme, we showed that when $p = 10^{-4}$, only 68 and 231 qubits are required to prepare an encoded $\ket{H}$ state with logical error rates $3.6 \times 10^{-8}$ and $4.9 \times 10^{-10}$ respectively. In addition, we also showed how our scheme can be used with encoded stabilizer operations to achieve significantly lower logical failure rates, both in the regime where $p = 10^{-4}$ and $p = 10^{-3}$. We stress that our results were obtained by considering a full circuit-level depolarizing noise model, where all stabilizer operations could fail.

We also point out key differences between magic state distillation schemes and our fault-tolerant methods for preparing magic states. Magic state distillation is a \textbf{top-down} approach, where error detection circuits are designed to prepare magic states without regard to their fault-tolerant properties. It is assumed that appropriate code distances for the encoded Clifford operations will be chosen to ensure that consecutive rounds of distillation will produce higher fidelity magic states. Our scheme, which uses flag-qubits and redundant ancilla encoding, is a \textbf{bottom-up} approach. The fault-tolerant properties of the magic state preparation circuits are prioritized and emphasis is given towards constructing $v$-flag circuits for large $v$. As such, to obtain magic states with very low failure rates, if encoded stabilizer operations are required, such operations can fail with error rates commensurate to the magic states being injected. 

The underlying codes that are used for our work belong to the triangular color code family. One avenue of exploration would be to consider the 4.8.8 color code family (see for instance Refs.~\cite{Landahl2011,TwistColorCode18}) for potentially better performance. In addition, the color codes used to encode the Clifford operations required two and three ancillas for the weight-four and weight-six stabilizers respectively. Using similar edge weight renormalization schemes to those described in Ref.~\cite{CKYZ20}, one could use fewer ancillas for measuring each stabilizer while maintaining the full effective code distance of the \texttt{Lift} decoder. Due to the smaller number of fault locations and reduced ancilla requirements, such an implementation could potentially significantly reduce the overhead for preparing encoded $\ket{H}$ states. 

When considering the implementation of our scheme with encoded stabilizer operations using the surface code, the $d_f = 7$ version of our scheme was optimal for both $p = 10^{-4}$ and $p = 10^{-3}$. A clear direction of future work would be to find a $v$-flag circuit (with $v \ge 4$) allowing a fault-tolerant implementation of a $d_f \ge 9$ scheme. Such a scheme could potentially further reduce the overhead for preparing $\ket{H}$ states with very low error rates. 

\begin{figure*}
\centering
\includegraphics[width=1.00\textwidth]{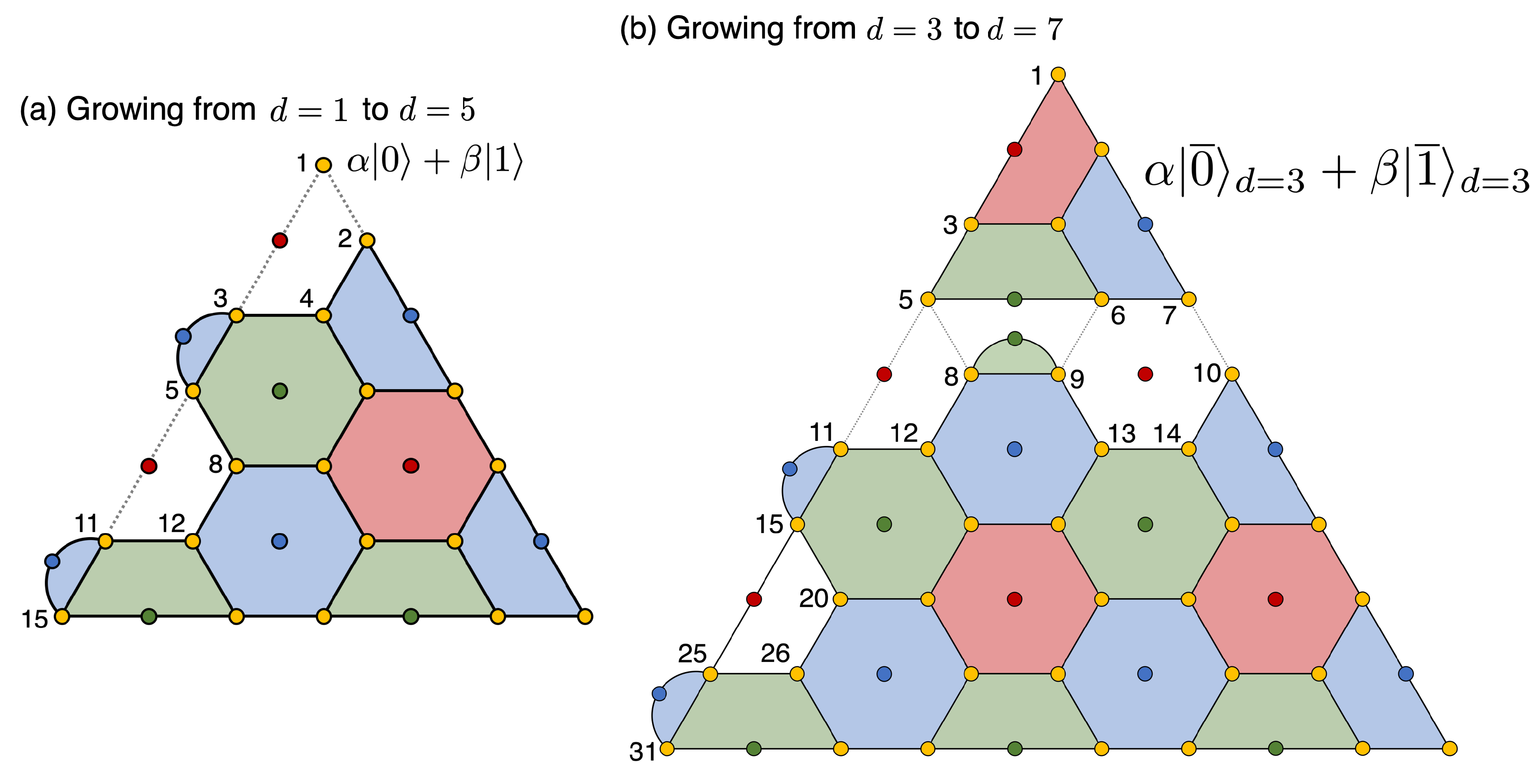}
\caption{Layouts for growing schemes used in the non fault-tolerant preparation of an encoded magic state. In (a), an input physical state $\alpha|0\rangle + \beta |1\rangle$ is grown to a logical state $\alpha|\overline{0}\rangle_{d=5} + \beta |\overline{1}\rangle_{d=5}$ encoded in the distance-$5$ triangular color code (see also \cref{fig:GrowingPhysical}). In (b), an input logical $\alpha|\overline{0}\rangle_{d=3} + \beta |\overline{1}\rangle_{d=3}$ encoded in the distance-$3$ triangular color code is grown to a logical state $\alpha|\overline{0}\rangle_{d=7} + \beta |\overline{1}\rangle_{d=7}$ encoded in the distance-$7$ triangular color code (see also \cref{fig:GeneralGrowingScheme}). }  
\label{fig:GrowingAppendix}
\end{figure*}

The schemes considered in this work to prepare $\ket{H}$ states are error detection schemes. In particular, for $p = 10^{-3}$ and $d_f > 3$, the acceptance probability for preparing an $\ket{H}$ state is very low (for instance, only $12 \%$ when $d_f = 5$). One way to improve the acceptance probability could be to use qubits encoded in a bosonic code \cite{AlbertNoh18} (such as a GKP code \cite{GottesmanGKP01}) and concatenate such qubits with the color code (the GKP code concatenated with the surface code was considered in Refs.~\cite{FTOF18,VuillotGKP19,NC20} for quantum memories). By using bosonic qubits, repeated rounds of error correction at the bosonic level prior to measuring the logical Hadamard operator and stabilizers of the color code could be performed to reduce some of the errors afflicting the data and ancilla qubits. Another possibility would be to develop an error correction scheme for preparing an $\ket{H}$ state which applies directly to the color code family. Such a scheme would have higher logical error rates compared to an error detection scheme, and the scheduling of the controlled-Hadamard gates would have to be considered more carefully. However, since an error correction scheme would not require any post selection, there could be an interval of physical error rates where it achieved better performance compared to the error detection scheme considered in this work. 

With a plethora of possible research directions for preparing magic states in a fault-tolerant way which build upon our work, and given the realistic hardware constraints that are built directly into our schemes, we believe our work paves the way for achieving very low overhead universal quantum computation with both near-term and long-term quantum devices.

\section*{Acknowledgments}
C.C. acknowledges Yihe Tang and Milan Cvitkovic for their help in setting up the computing resources with the AWS clusters which were used for performing all the numerical simulations in this work. We would like to thank Earl Campbell, Aleksander Kubica and Fernando Brandao for useful discussions. We thank Simone Severini, Bill Vass and Dominique L'Eplattenier for their guidance and help with submitting the paper. We also thank Kevin Dothager for his help with IP.

\appendix

\section{Non fault-tolerant $\ket{H}$ state preparation scheme}
\label{appendix:NonFaultTolerantPrepScheme}

In this section, we show how measuring the white plaquettes when implementing the growing schemes in \cref{fig:GrowingPhysical,fig:GeneralGrowingScheme} and applying the corrections from the matching graphs results in the correct encoded state. Let us first consider the growing scheme shown in \cref{fig:GrowingAppendix}a that converts a physical input state $\alpha|0\rangle + \beta |1\rangle$ into a logical state $\alpha|\overline{0}\rangle_{d=5} + \beta |\overline{1}\rangle_{d=5}$ encoded in the distance-$5$ triangular color code. In this scheme, we first prepare a stabilizer state $|S_{t}\rangle$ that is stabilized by $14$ out of the $18$ stabilizer generators of the distance-$5$ triangular color code  
\begin{align}
g_{X}^{(3)},\cdots,g_{X}^{(9)} , \,\,\,\textrm{and}\,\,\, g_{Z}^{(3)},\cdots,g_{Z}^{(9)}, \label{eq:stabilizers from physical to logical from color code} 
\end{align}
as well as the following four weight-2 stabilizers 
\begin{alignat}{2}
g'^{(1)}_{X} &= X_{3}X_{5}, & \quad g'^{(2)}_{X} &= X_{11}X_{15},
\nonumber\\
g'^{(1)}_{Z} &= Z_{3}Z_{5}, & \quad g'^{(2)}_{Z} &= Z_{11}Z_{15}. \label{eq:stabilizers from physical to logical weight 2}
\end{alignat}
We define $\mathcal{S}_{\text{st}}$ to be the group generated by the operators in \cref{eq:stabilizers from physical to logical from color code,eq:stabilizers from physical to logical weight 2}. Note that $\ket{S_t}$ can be prepared by first preparing all qubits in the $\ket{0}$ state, and then measuring only the $X$-type generators in $\mathcal{S}_{\text{st}}$. As such, we have 
\begin{align}
|S_{t}\rangle &\propto \Big{[} \prod_{k=1}^{2}(I+g'^{(k)}_{X}) \Big{]} \Big{[} \prod_{k=3}^{9} (I+g_{X}^{(k)}) \Big{]} |0\rangle^{\otimes18}. 
\end{align}  
One can readily check that $|S_{t}\rangle$ is stabilized by all the $18$ stabilizers in \cref{eq:stabilizers from physical to logical from color code,eq:stabilizers from physical to logical weight 2} as desired.

Once $\ket{S_t}$ is prepared, we measure the four missing stabilizer generators of the distance-$5$ triangular code given by
\begin{alignat}{2}
g^{(1)}_{X} &= X_{1}X_{2}X_{3}X_{4}, & \quad g^{(2)}_{X} &= X_{5}X_{8}X_{11}X_{12},
\nonumber\\
g^{(1)}_{Z} &= Z_{1}Z_{2}Z_{3}Z_{4}, & \quad g^{(2)}_{Z} &= Z_{5}Z_{8}Z_{11}Z_{12}. 
\label{eq:White11}
\end{alignat} 
Such operators are the white plaquettes shown in \cref{fig:GrowingAppendix}a. We define $\mathcal{S}_{b_1}$ to the group generated by the operators in \cref{eq:White11}.

Initially, the system is in the state 
\begin{align}
|\psi_{0}\rangle = (\alpha|0\rangle + \beta|1\rangle) \otimes |S_{t}\rangle  . 
\end{align}
After measuring the $X$-type stabilizers of $\mathcal{S}_{b_1}$, we get the following state
\begin{align}
|\psi_{1}^{[m_{1}m_{2}]}\rangle &\propto \prod_{k=1}^{2} (I+(-1)^{m_{k}}g^{(k)}_{X}) |\psi_{0}\rangle, 
\end{align} 
conditioned on obtaining $( (-1)^{m_{1}} ,  (-1)^{m_{2}})$ where $m_{1},m_{2}\in\lbrace  0,1\rbrace$ when measuring $g^{(1)}_{X}$ and $g^{(2)}_{X}$. In any case, we can always convert the state $|\psi_{1}^{[m_{1}m_{2}]}\rangle$ to $|\psi_{1}^{[00]}\rangle$ by applying a correction operator 
\begin{align}
\begin{cases}
I & (m_{1},m_{2}) =(0,0) \\
g'^{(2)}_{Z} & (m_{1},m_{2}) =(0,1) \\
g'^{(1)}_{Z}g'^{(2)}_{Z} & (m_{1},m_{2}) =(1,0) \\
g'^{(1)}_{Z} & (m_{1},m_{2}) =(1,1) 
\end{cases} \label{eq:correction operator from physical to distance 5}
\end{align} 
since $g'^{(1)}_{Z}$ anti-commutes with $g_{X}^{(1)}$ and $g_{X}^{(2)}$, and $g'^{(2)}_{Z}$ anti-commutes with $g_{X}^{(2)}$. Note that the correction operators in \cref{eq:correction operator from physical to distance 5} can be determined by implementing MWPM on the matching graph $G^{(5)}_{1x}$ shown in \cref{fig:GrowingPhysical}b. Thus after the correction, we are always left with the state 
\begin{align}
&|\psi_{1}^{[00]}\rangle  \propto \prod_{k=1}^{2} (I+g^{(k)}_{X}) |\psi_{0}\rangle
\nonumber\\
&\propto (\alpha I +\beta X_{1})  \Big{[} \prod_{k=1}^{2}(I+g'^{(k)}_{X}) \Big{]} \Big{[} \prod_{k=1}^{9} (I+g_{X}^{(k)}) \Big{]} |0\rangle^{19}, 
\end{align}  
where we used $\alpha|0\rangle + \beta |1\rangle = (\alpha I + \beta X_{1})|0\rangle$.    

Next, we measure the two $Z$-type stabilizers of $\mathcal{S}_{b_1}$. After the measurement, the state becomes
\begin{align}
|\psi_{2}^{[m_{1}m_{2}]}\rangle &\propto \prod_{k=1}^{2} (I+(-1)^{m_{k}}g^{(k)}_{Z}) |\psi_{1}^{[00]}\rangle, 
\end{align}
conditioned on the measurement outcomes $(  (-1)^{m_{1}} ,  (-1)^{m_{2}})$. Performing the same steps as above, the state $|\psi_{2}^{[m_{1}m_{2}]}\rangle$ can be converted to $|\psi_{2}^{[00]}\rangle$ by applying an appropriate correction operator which is determined by implementing MWPM on the matching graph $G^{(5)}_{1z}$ shown in \cref{fig:GrowingPhysical}b. Afterwords, we are left with 
\begin{align}
&|\psi_{2}^{[00]}\rangle \propto \prod_{k=1}^{2} (I+g^{(k)}_{Z}) |\psi_{1}^{[00]}\rangle
\nonumber\\
&\propto \Big{[} \prod_{k=1}^{2}(I+g^{(k)}_{Z}) \Big{]}  (\alpha I +\beta X_{1})  \Big{[} \prod_{k=1}^{2}(I+g'^{(k)}_{X}) \Big{]} 
\nonumber\\
&\quad \times \Big{[} \prod_{k=1}^{9} (I+g_{X}^{(k)}) \Big{]} |0\rangle^{19}
\nonumber\\
&\propto \Big{[} \prod_{k=1}^{2}(I+g^{(k)}_{Z}) \Big{]}  (\alpha I +\beta X_{1})  \Big{[} \prod_{k=1}^{2}(I+g'^{(k)}_{X}) \Big{]}  |\overline{0}\rangle_{d=5}. 
\end{align}
where $|\overline{0}\rangle_{d=5}$ is the logical zero state of the distance $5$ triangular color code. Note that 
\begin{align}
&\Big{[} \prod_{k=1}^{2}(I+g^{(k)}_{Z}) \Big{]}  (\alpha I +\beta X_{1})  \Big{[} \prod_{k=1}^{2}(I+g'^{(k)}_{X}) \Big{]}  |\overline{0}\rangle_{d=5}
\nonumber\\
&= \alpha I (I+g^{(1)}_{Z}) (I+g'^{(1)}_{X})  (I+g^{(2)}_{Z})  (I+g'^{(2)}_{X}) |\overline{0}\rangle_{d=5}
\nonumber\\
&\quad +  \beta X_{1}  (I-g^{(1)}_{Z})  (I+g'^{(1)}_{X})  (I+g^{(2)}_{Z}) (I+g'^{(2)}_{X}) |\overline{0}\rangle_{d=5}
\nonumber\\
&=  \alpha I  (I+g^{(2)}_{Z})   (I+g'^{(2)}_{X})  (I+g^{(1)}_{Z}) (I+g'^{(1)}_{X}) |\overline{0}\rangle_{d=5}
\nonumber\\
&\quad +  \beta X_{1}  (I+g^{(2)}_{Z})  (I+g'^{(2)}_{X}) (I-g^{(1)}_{Z})   (I+g'^{(1)}_{X})  |\overline{0}\rangle_{d=5}
\nonumber\\
&\propto  \alpha I  (I+g^{(2)}_{Z})   (I+g'^{(2)}_{X})  |\overline{0}\rangle_{d=5}
\nonumber\\
&\quad +  \beta X_{1}  (I+g^{(2)}_{Z})  (I+g'^{(2)}_{X}) g'^{(1)}_{X}  |\overline{0}\rangle_{d=5}
\nonumber\\
&=  \alpha I  (I+g^{(2)}_{Z})   (I+g'^{(2)}_{X})  |\overline{0}\rangle_{d=5}
\nonumber\\
&\quad +  \beta X_{1} g'^{(1)}_{X}  (I-g^{(2)}_{Z})  (I+g'^{(2)}_{X})  |\overline{0}\rangle_{d=5}
\nonumber\\
&\propto ( \alpha I  +  \beta X_{1} g'^{(1)}_{X}  g'^{(2)}_{X} )  |\overline{0}\rangle_{d=5}
\nonumber\\
&\propto  ( \alpha I  +  \beta X_{1}X_{3}X_{5}X_{11}X_{15} )  |\overline{0}\rangle_{d=5}. \label{eq:reasoning from physical to 5 growing scheme} 
\end{align} 
Since $\overline{X} = X_{1}X_{3}X_{5}X_{11}X_{15}$ is the logical X operator of the distance-$5$ triangular color code, we can conclude that the output state $|\psi_{2}^{[00]}\rangle$ is given by
\begin{align}
|\psi_{2}^{[00]}\rangle &\propto ( \alpha I  +  \beta \overline{X} )  |\overline{0}\rangle_{d=5} = \alpha |\overline{0}\rangle_{d=5} + \beta |\overline{1}\rangle_{d=5}. 
\end{align}
Hence, the output state is the is the desired state $\alpha \ket{\overline{0}} + \beta \ket{\overline{1}}$ encoded in the $d=5$ triangular color code. 

We now move on to the growing scheme shown in \cref{fig:GrowingAppendix}b that converts an input state $\alpha|\overline{0}\rangle_{d=3}+\beta|\overline{1}\rangle_{d=3}$ (encoded in the $d=3$ triangular color code) into a logical state $\alpha|\overline{0}\rangle_{d=7}+\beta|\overline{1}\rangle_{d=7}$ encoded in the $d=7$ triangular color code. As in the previous scheme, we first prepare a stabilizer state $|S_{t}\rangle$ that is stabilized by $24$ out of the $36$ generators of the $d=7$ triangular color code as well as $6$ weight-$2$ stabilizers which are given by
\begin{alignat}{4}
g'^{(1)}_{X} &= X_{8}X_{9}, & \quad g'^{(2)}_{X} &= X_{11}X_{15} &\quad g'^{(3)}_{X} &= X_{25}X_{31},
\nonumber\\
g'^{(1)}_{Z} &= Z_{8}Z_{9}, & \quad g'^{(2)}_{Z} &= Z_{11}Z_{15} &\quad g'^{(3)}_{Z} &= Z_{25}Z_{31}. 
\end{alignat}
To initiate the growing scheme, we measure the following $6$ stabilizers of the $d=7$ triangular color code, which are represented by the white plaquettes of \cref{fig:GrowingAppendix}b
\begin{align}
g^{(1)}_{X} &= X_{6}X_{7}X_{9}X_{10}X_{13}X_{14}  
\nonumber\\
g^{(2)}_{X} &= X_{5}X_{8}X_{11}X_{12}, \quad  g^{(3)}_{X} = X_{15}X_{20}X_{25}X_{26},
\nonumber\\
g^{(1)}_{Z} &= Z_{6}Z_{7}Z_{9}Z_{10}Z_{13}Z_{14}  
\nonumber\\
g^{(2)}_{Z} &= Z_{5}Z_{8}Z_{11}Z_{12}, \quad  g^{(3)}_{Z} = Z_{15}Z_{20}Z_{25}Z_{26}. \label{eq:missing color code stabilizers in the 3 to 7 growing scheme}
\end{align}
Note that the operators in \cref{eq:missing color code stabilizers in the 3 to 7 growing scheme} are not stabilizers of the stabilizer state $|S_{t}\rangle$ and the $d=3$ triangular color code (which is being merged with $|S_{t}\rangle$).

Once the stabilizer state is prepared, and prior to measuring the operators in \cref{eq:missing color code stabilizers in the 3 to 7 growing scheme}, the system is in the state
\begin{align}
|\psi_{0}\rangle &=  (\alpha |\overline{0}\rangle_{d=3} + \beta |\overline{1}\rangle_{d=3} ) \otimes |S_{t}\rangle  . 
\end{align}
After measuring the three $X$-type stabilizers in \cref{eq:missing color code stabilizers in the 3 to 7 growing scheme}, we get the following state 
\begin{align}
|\psi_{1}^{[m_{1}m_{2}m_{3}]}\rangle &\propto \prod_{k=1}^{3} ( I + (-1)^{m_{k}} g^{(k)}_{X}  ) |\psi_{0}\rangle, 
\end{align} 
where the values of $m_1,m_2,m_3 \in \{ 0, 1 \}$ depend on the measurement outcomes of $g^{(1)}_{X}$,  $g^{(2)}_{X}$ and $g^{(3)}_{X}$. As in the case where a physical state was grown to an encoded state of the $d=5$ triangular color code, we can convert any output state $|\psi_{1}^{[m_{1}m_{2}m_{3}]}\rangle$ to $|\psi_{1}^{[000]}\rangle$ by applying a correction operator 
\begin{align}
\begin{cases}
I & (m_{1},m_{2},m_{3}) = (0,0,0) \\
g'^{(3)}_{Z} & (m_{1},m_{2},m_{3}) = (0,0,1) \\
g'^{(2)}_{Z}g'^{(3)}_{Z} & (m_{1},m_{2},m_{3}) = (0,1,0) \\
g'^{(2)}_{Z} & (m_{1},m_{2},m_{3}) = (0,1,1) \\
g'^{(1)}_{Z}g'^{(2)}_{Z}g'^{(3)}_{Z} & (m_{1},m_{2},m_{3}) = (1,0,0) \\
g'^{(1)}_{Z}g'^{(2)}_{Z} & (m_{1},m_{2},m_{3}) = (1,0,1) \\
g'^{(1)}_{Z} & (m_{1},m_{2},m_{3}) = (1,1,0) \\
g'^{(1)}_{Z}g'^{(3)}_{Z} & (m_{1},m_{2},m_{3}) = (1,1,1) 
\end{cases} . \label{eq:correction operator 3 to 7 growing scheme}
\end{align}
Note that correction operators of \cref{eq:correction operator 3 to 7 growing scheme} can be determined by implementing MWPM on the matching graph $G^{(7)}_{3x}$ shown in \cref{fig:GeneralGrowingScheme}b. After the applying such corrections, we are always left with the state 
\begin{align}
|\psi_{1}^{[000]}\rangle &\propto \prod_{k=1}^{3} ( I + g^{(k)}_{X}  ) |\psi_{0}\rangle. 
\label{eq:IntStateBefZ}
\end{align} 

The final step consists of measuring the three $Z$-type stabilizers in \cref{eq:missing color code stabilizers in the 3 to 7 growing scheme}. The state in \cref{eq:IntStateBefZ} then becomes
\begin{align}
\ket{\psi_{2}^{[m_{1}m_{2}m_{3}]}} &\propto \prod_{k=1}^{3} ( I + (-1)^{m_{k}} g^{(k)}_{Z}  ) |\psi_{1}^{[000]}\rangle, 
\end{align} 
where the values of $m_1,m_2,m_3 \in \{ 0, 1 \}$ depend on the measurement outcomes of $g^{(1)}_{Z}$,  $g^{(2)}_{Z}$ and $g^{(3)}_{Z}$. The states $\ket{\psi_{2}^{[m_{1}m_{2}m_{3}]}}$ can be mapped to the state $|\psi_{2}^{[000]}\rangle$ again by applying an appropriate correction operator as was done in \cref{eq:correction operator 3 to 7 growing scheme}, but with $X$-type operators. Thus, at the end of the growing scheme, we have 
\begin{align}
|\psi_{2}^{[000]}\rangle &\propto  \Big{[} \prod_{k=1}^{3} ( I + g^{(k)}_{Z}  ) \Big{]} |\psi_{1}^{[000]}\rangle,
\nonumber\\
&\propto  \Big{[} \prod_{k=1}^{3} ( I + g^{(k)}_{Z}  ) \Big{]}\Big{[} \prod_{k=1}^{3} ( I + g^{(k)}_{X}  ) \Big{]} |\psi_{0}\rangle. 
\end{align} 
Repeating a similar analysis as in \cref{eq:reasoning from physical to 5 growing scheme}, we get the following desired result 
\begin{align}
|\psi_{2}^{[000]}\rangle &\propto \alpha |\overline{0}\rangle_{d=7} + \beta |\overline{1}\rangle_{d=7} . 
\end{align}   

\section{Proof of fault-tolerance for the $\ket{\overline{H}}$ state preparation scheme in \cref{sec:FaultTolHstate}.}
\label{appendix:FaultFreeHadMeas}

In this section we show that in order for the magic state preparation protocol of \cref{sec:FaultTolHstate} to be fault-tolerant, the pair of $H^{(d)}_m$ and $EC^{(d)}$ circuits need to be repeated a minimum of $ (d-1)/2$ times. In what follows, we say that a state preparation protocol is $t$-fault-tolerant if the following two conditions are satisfied (see for instance Refs.\cite{Gottesman2010,ChamberlandMagic}):

\begin{definition}{\underline{Fault-tolerant state preparation}}
	
	For $t = (d-1)/2$, a state-preparation protocol using a distance-$d$ stabilizer code $C$ is $t$-fault-tolerant if the following two conditions are satisfied:
	\begin{enumerate}
		\item If there are $s$ faults during the state-preparation protocol with $s \le t$, the resulting state differs from a codeword by an error of at most weight $s$.
		\item If there are $s$ faults during the state-preparation protocol with $s \le t$, then ideally decoding the output state results in the same state that would be obtained from the fault-free state-preparation scheme. 
	\end{enumerate}
	\label{Def:FaultTolerantPrep}
\end{definition}
Here ideally decoding refers to performing a round of fault-free error correction. In what follows, the code $C$ belongs to the triangular color code family. Further, since the $H^{(d)}_m$ circuits are valid for $d \in \{3,5,7\}$, the following arguments apply for code distances of the triangular color code that are no greater than seven. 

We have already verified numerically that the $H^{(d)}_m$ and $EC^{(d)}$ circuits are $t$-flag circuits (with $t = (d-1)/2$). Hence if there are $s \le t$ faults in the $H^{(d)}_m$ or $EC^{(d)}$ circuits resulting in an error $E$ with either $\text{min}(\text{wt}(E), \text{wt}(E\overline{H}))  > s$ or  $\text{min}(\text{wt}(E), \text{wt}(EP))  > s$ (for any of the stabilizers $P$), at least one flag qubit will flag and the protocol aborts. 

\begin{figure}
	\centering
	\includegraphics[width=0.4\textwidth]{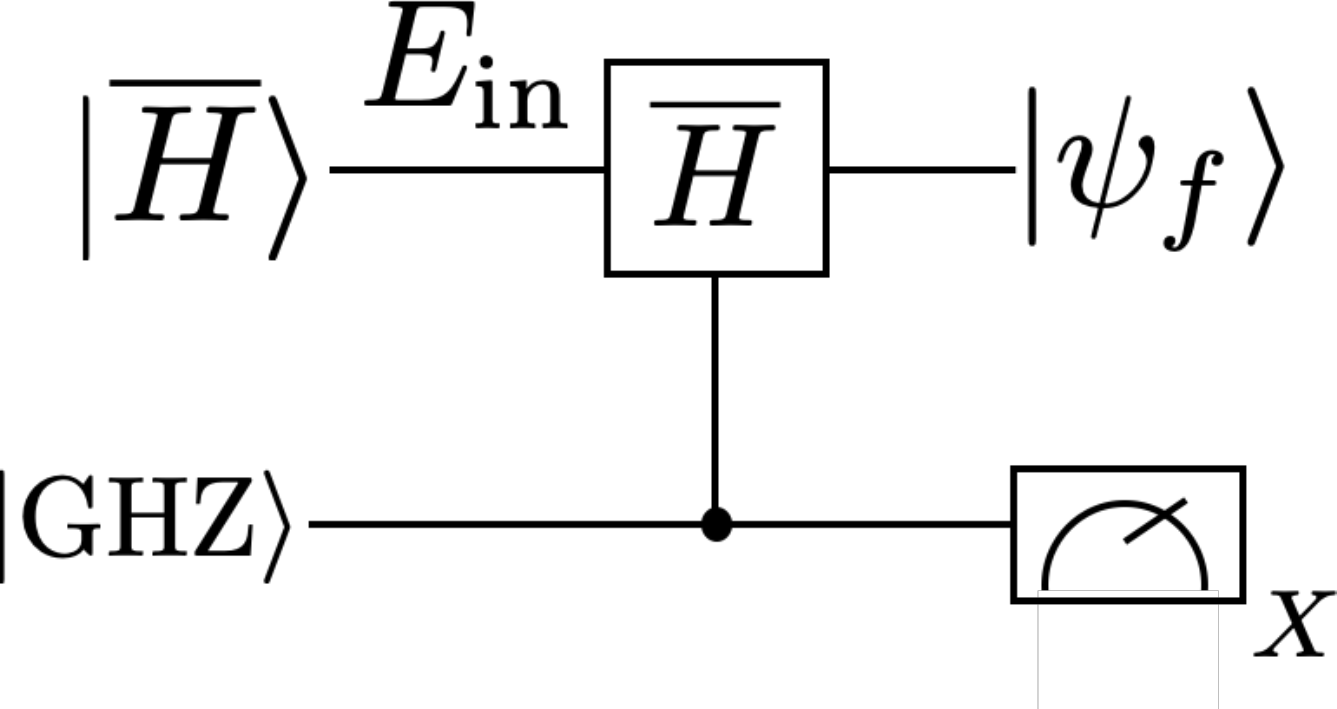}
	\caption{Schematic for the implementation of the $H^{(d)}_{m}$ circuit with an input error $E_{\text{in}}$, resulting in an output state $\ket{\psi_f}$. A GHZ state is prepared and the parity of the logical Hadamard operator $\overline{H} = H^{\otimes n}$ is measured. We omit the details for the fault-tolerant preparation of the GHZ state (see for instance \cref{fig:MeasureHd3Detailed}) as it is not important for the discussion in this section.}
	\label{fig:HadMeasErrTrans}
\end{figure}

First, we consider all the possible output errors of the fault-free implementation of the $H^{(d)}_m$ circuits given an input error $E_{\text{in}}$ (see \cref{fig:HadMeasErrTrans}). The input error $E_{\text{in}}$ arises from faults which occur during the implementation of the $G^{(1 \rightarrow d)}$ circuit (see for instance \cref{fig:GeneralHstatePrepScheme,fig:GrowingPhysical}). We illustrate all possible cases for $E_{\text{in}}$ and for each case we show the resulting output errors. Note that in what follows, the state $\ket{\text{GHZ}} = \frac{1}{\sqrt{2}}( |0\rangle^{\otimes m} +|1\rangle^{\otimes m} )$ (where $m$ is the number of ancilla qubits) can be replaced by the single-qubit $\ket{+}$ state without changing the final result. We also write the controlled-Hadamard gate as $C_{\overline{H}}$. Lastly, given an error $E$, $s(E)$ will correspond to the error syndrome of $E$ obtained by measuring all the stabilizer generators of the underlying stabilizer code used to encode the data. 

\underline{Case 1: $E_{\text{in}} = \overline{X}$ or $E_{\text{in}} = \overline{Z}$}.

Suppose  $E_{\text{in}} = \overline{X}$. Prior to performing the $C_{\overline{H}}$ gate, we have
\begin{align}
\ket{\psi_1} &= \overline{X}\ket{\overline{H}}\otimes \ket{+} \nonumber \\
&= \frac{1}{\sqrt{2}}(\overline{X}\ket{\overline{H}}\otimes \ket{0} + \overline{X}\ket{\overline{H}}\otimes \ket{1}).
\end{align}
Applying the $C_{\overline{H}}$ gate and using the identity $HX = ZH$, $\ket{\psi_1}$ transforms to $\ket{\psi_2}$ which is given by
\begin{align}
\ket{\psi_2}  &=  \frac{1}{\sqrt{2}}(\overline{X}\ket{\overline{H}}\otimes \ket{0} + \overline{Z}\ket{\overline{H}}\otimes \ket{1}) \nonumber \\
& = \frac{1}{\sqrt{2}}(\frac{\overline{X} + \overline{Z}}{\sqrt{2}}\ket{\overline{H}}\otimes \ket{+} + \frac{\overline{X} - \overline{Z}}{\sqrt{2}}\ket{\overline{H}}\otimes \ket{-}) \nonumber \\
&=  \frac{1}{\sqrt{2}}(\ket{\overline{H}}\otimes \ket{+} + \frac{\overline{X} - \overline{Z}}{\sqrt{2}}\ket{\overline{H}}\otimes \ket{-}).
\label{eq:Ein11}
\end{align}

Similarly, if $E_{\text{in}} = \overline{Z}$, performing the same steps shows that 
\begin{align}
\ket{\psi_2} = \frac{1}{\sqrt{2}}(\ket{\overline{H}}\otimes \ket{+} - \frac{\overline{X} - \overline{Z}}{\sqrt{2}}\ket{\overline{H}}\otimes \ket{-}).
\label{eq:Ein12}
\end{align}

From \cref{eq:Ein11,eq:Ein12}, we see that performing the $X$-basis measurement and discarding the ancilla, the final output state $\ket{\psi_3}$ is $\ket{\psi_f} = \ket{\overline{H}}$ if a $+1$ outcome is obtained (which is the desired state), and $\ket{\psi_f} = \frac{\overline{X} - \overline{Z}}{\sqrt{2}}\ket{\overline{H}}$ if a $-1$ outcome is obtained (each occurring with a $50 \%$ probability).

\underline{Case 2: $E_{\text{in}} = \overline{Y}$}.

Prior to performing the $C_{\overline{H}}$ gate, we have
\begin{align}
\ket{\psi_1} &= \overline{Y}\ket{\overline{H}}\otimes \ket{+} \nonumber \\
&= \frac{1}{\sqrt{2}}(\overline{Y}\ket{\overline{H}}\otimes \ket{0} + \overline{Y}\ket{\overline{H}}\otimes \ket{1}).
\end{align}
Applying the  $C_{\overline{H}}$ gate and using the identity $HY = -YH$, we have 
\begin{align}
\ket{\psi_2} &= \frac{1}{\sqrt{2}}(\overline{Y}\ket{\overline{H}}\otimes \ket{0} - \overline{Y}\ket{\overline{H}}\otimes \ket{1}) \nonumber \\
&= \overline{Y}\ket{\overline{H}} \otimes \ket{-}.
\label{eq:Eq21}
\end{align}
Hence the ancilla measurement outcome will always be $-1$ with the final output state $\ket{\psi_f} = \overline{Y}\ket{\overline{H}}$. 

\underline{Case 3: $E_{\text{in}} = E'\overline{Y}$}.

In this case, we assume that $s(E') \neq \textbf{0}$ where $\textbf{0}$ is the all zeros bit string of length $n-1$ (where $n$ is the number of data qubits of the underlying stabilizer code encoding the data). Further, we define $\tilde{E}' = \overline{H} E' \overline{H}^{\dagger}$. Performing an analogous calculation to the one leading to \cref{eq:Eq21}, we have
\begin{align}
\ket{\psi_2} = \frac{1}{\sqrt{2}}(\frac{(E' - \tilde{E}')}{\sqrt{2}} \overline{Y}\ket{\overline{H}} \otimes \ket{+} + \frac{(E' + \tilde{E}')}{\sqrt{2}} \overline{Y}\ket{\overline{H}} \otimes \ket{-}).
\end{align}
Hence, if the ancilla is measured as $+1$, the output state will be $\ket{\psi_f} = \frac{(E' - \tilde{E}')}{\sqrt{2}} \overline{Y}\ket{\overline{H}}$ whereas a $-1$ outcome will yield $\ket{\psi_f} = \frac{(E' + \tilde{E}')}{\sqrt{2}} \overline{Y}\ket{\overline{H}}$. Both measurement outcomes occur with $50 \%$ probability. Further, note that $\frac{(E' - \tilde{E}')}{\sqrt{2}}$ and $\frac{(E' + \tilde{E}')}{\sqrt{2}}$ are detectable errors.

\underline{Case 4: $E_{\text{in}} = E'\overline{X}$ or $E_{\text{in}} = E'\overline{Z}$}.

Again, we assume that $s(E') \neq \textbf{0}$. Performing the same calculations as above, we find
\begin{align}
\ket{\psi_2} &= \frac{1}{\sqrt{2}}(\frac{(E'\overline{X} + \tilde{E}'\overline{Z})}{\sqrt{2}} \ket{\overline{H}} \otimes \ket{+} \nonumber \\
&+ \frac{(E'\overline{X} - \tilde{E}'\overline{Z})}{\sqrt{2}}\ket{\overline{H}} \otimes \ket{-}).
\label{eq:ImportantProof}
\end{align}
Again, the ancilla measurement outcomes will be $\pm1$, each occurring with $50 \%$ probability. A $+1$ outcome yields the output state $\ket{\psi_f} = \frac{(E'\overline{X} + \tilde{E}'\overline{Z})}{\sqrt{2}} \ket{\overline{H}}$ and a $-1$ outcome yields $\ket{\psi_f} = \frac{(E'\overline{X} - \tilde{E}'\overline{Z})}{\sqrt{2}} \ket{\overline{H}}$. In both cases, the errors afflicting the state $\ket{\overline{H}}$ will be detected by a fault-free $EC^{(d)}$ circuit. The case where $E_{\text{in}} = E'\overline{Z}$ yields identical output states, up to a global sign. We will now explain why the $H^{(d)}_m$ and $EC^{(d)}$ circuits need to come in pairs.

In the example provided leading to \cref{eq:ImportantProof}, if the $+1$ measurement outcome is obtained, there can be an fault resulting in the error $E'$ at the very beginning of the subsequent $EC^{(d)}$ circuit cancelling the term multiplying $\overline{X}$ and thus resulting in the error $E=\frac{\overline{X} + E'\tilde{E}'\overline{Z}}{\sqrt{2}}$. As such, there is a $50 \%$ chance that a trivial syndrome is obtained when implementing the $EC^{(d)}$ circuit, resulting in the output error $\overline{X}$. However, by applying the $H^{(d)}_m$ circuit a second time (assuming it is fault-free), Case 1 shows that a $+1$ outcome cannot result in an output state with a logical fault. Since the pair of $H^{(d)}_m$ and $EC^{(d)}$ circuits are repeated $(d-1)/2$ times, at least one such pair must be fault-free if the total number of faults $v$ has $v \le (d-1)/2$. This example illustrates the importance of applying the $H^{(d)}_m$ and $EC^{(d)}$ circuits in pairs. For instance, if all the $H^{(d)}_m$ circuits were repeated $(d-1)/2$ times, followed by the repetition of the $EC^{(d)}$ circuits $(d-1)/2$ times, an error of the form $\frac{(E'\overline{X} + \tilde{E}'\overline{Z})}{\sqrt{2}}$ would always result in a $+1$ outcome of the $H^{(d)}_m$ circuits and the output error would be unchanged. Then as shown above, a single fault at the beginning of the first $EC^{(d)}$ circuit could result in a trivial syndrome with $50 \%$ probability (with an output error $\overline{X}$) and all subsequent rounds of syndrome measurement would yield the trivial syndrome. 

\begin{figure*}
\centering
\includegraphics[width=0.8\textwidth]{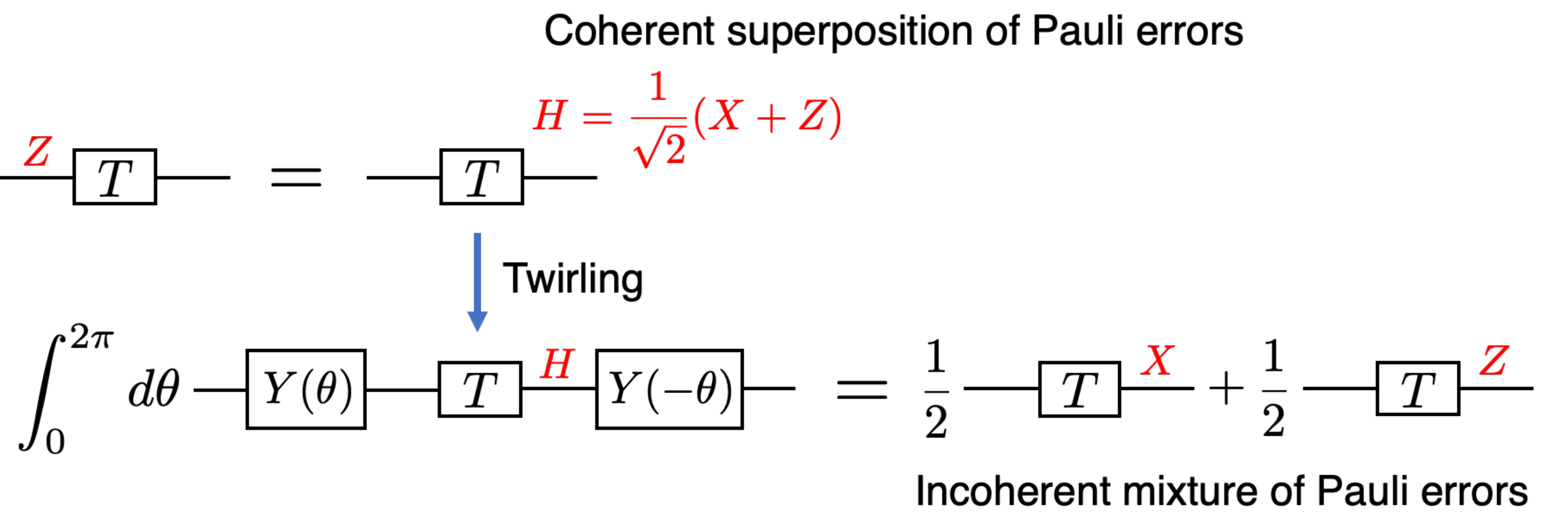}
\caption{Conversion of a Hadamard error $H = (X+Z)/\sqrt{2}$ (i.e., coherent superposition of Pauli errors) into an incoherent mixture of Pauli errors $X$ and $Z$ via a noise twirling. Note that $Y(\theta)$ is defined as $Y(\theta) \equiv \exp[ -i (\theta/2) Y ]$ and $T= Y(\pi/4)$. }
\label{fig:TgateErrorTwirling}
\end{figure*}

Now suppose there are $s = (d-1)/2$ faults spread throughout all the circuits illustrated in \cref{fig:GeneralHstatePrepScheme}. We consider the worst case scenario, where a single fault results in an error $E$ (which should be detected by our protocol), and the remaining $s-1$ faults all result in measurement errors (and such faults can potentially add additional data qubit errors, say if it arises from a CNOT gate) preventing $E$ from being detected in either the $H^{(d)}_m$ or $EC^{(d)}$ circuits. Since the pair of $H^{(d)}_m$ and $EC^{(d)}$ circuits are repeated $(d-1)/2$ times, at least one pair of $H^{(d)}_m$ and $EC^{(d)}$ circuits will be fault-free and thus not afflicted by measurement errors. From the above, if the $H^{(d)}_m$ and $EC^{(d)}$ circuits are fault-free, then $+1$ measurement outcomes in both circuits cannot yield an uncorrectable output error. Further, since all the $H^{(d)}_m$ and $EC^{(d)}$ circuits are $t$-flag circuits (with $t = (d-1)/2$), the final output error $E_{\text{final}}$ must have $\text{wt}(E_{\text{final}}) \le s$ and as such, it must be a correctable error. 

We note that in the general case, where a $\ket{\text{GHZ}}$ state is used instead of the $\ket{+}$ state, we can replace $+1$ and $-1$ measurement outcomes with even and odd parity measurement outcomes, and the same conclusions would follow. 

\section{Twirling approximation}
\label{appendix:TwirlingApprox}

In this section, we show that a non-Pauli error after a $T^{\dagger}$ or $T$ gate can be converted via a noise twirling operation into an incoherent mixture of Pauli errors. To be clear, we do not propose to physically perform the noise twirling after the $T^{\dagger}$ and $T$ gates as part of the protocol as this can reduce the performance of our scheme. Instead, we aim to show that the approximations we performed in our numerical simulations are justified. 

Suppose for instance there is an input $Z$ error to a $T$ gate. Recall that the input Pauli error $Z$ is converted through the $T$ gate into a non-Pauli error $H = \frac{1}{\sqrt{2}} (X+Z)$. We make this non-Pauli error into an incoherent mixture of Pauli errors by applying $Y(\theta)$ and $Y(-\theta)$ before and after the noisy $T$ gate where the rotation angle $\theta$ is drawn uniformly from the range $[0,2\pi]$. Note that 
\begin{align}
\rho'  &\equiv \int_{0}^{2\pi}d\theta Y(-\theta) H T Y(\theta) \rho Y(-\theta) T^{\dagger} H Y(\theta)
\nonumber\\
&= \int_{0}^{2\pi}d\theta Y(-\theta) H Y(\theta)  (T \hat{\rho} T^{\dagger}) Y(-\theta)  H Y(\theta), 
\end{align}
where we used the fact that $T=Y(\frac{\pi}{4})$ commutes with $Y(\theta)$ for any $\theta\in [0,2\pi]$. Then, since 
\begin{align}
Y(-\theta) X Y(\theta) &=  \cos\theta X + \sin\theta Z, 
\nonumber\\
Y(-\theta) Z Y(\theta) &=  - \sin\theta X + \cos\theta Z , 
\end{align}
we have 
\begin{align}
Y(-\theta) H  Y(\theta) &= \frac{1}{\sqrt{2}}Y(-\theta) (X+Z) Y(\theta)
\nonumber\\
&= \cos\Big{(}\theta + \frac{\pi}{4}\Big{)} X + \sin\Big{(}\theta + \frac{\pi}{4}\Big{)} Z, 
\end{align}
and thus 
\begin{align}
\rho' &= \int_{0}^{2\pi}d\theta \Big{[} \cos^{2}\Big{(}\theta + \frac{\pi}{4}\Big{)} X(T\rho T^{\dagger}) X 
\nonumber\\
&\qquad\qquad + \sin^{2}\Big{(}\theta + \frac{\pi}{4}\Big{)} Z(T\rho T^{\dagger}) Z 
\nonumber\\
&\qquad\qquad +\frac{1}{2}\sin\Big{(} 2\theta + \frac{\pi}{2} \Big{)} X(T\rho T^{\dagger}) Z  
\nonumber\\
&\qquad\qquad + \frac{1}{2}\sin\Big{(} 2\theta + \frac{\pi}{2} \Big{)} Z(T\rho T^{\dagger}) X   \Big{]}
\nonumber\\
&= \frac{1}{2}\Big{[} X(T\rho T^{\dagger})X + Z(T\rho T^{\dagger})Z  \Big{]} . 
\end{align} 
That is, the output Hadamard error $H = \frac{1}{\sqrt{2}}(X+Z)$ is converted via the noise twirling to an incoherent mixture of the Pauli $X$ and $Z$ errors, each with $50\%$ probability (see \cref{fig:TgateErrorTwirling}). The same reasoning holds for any output error $\cos\phi X + \sin\phi Z$ for any $\phi \in [0,2\pi]$. On the other hand, since a Pauli $Y$ error commutes with the $T$ gate and the $Y(\theta)$ gates, it is unaffected by the noise twirling and remains to be a Pauli $Y$ error.    

\bibliographystyle{unsrtnat} 
\bibliography{bibtex_chamberland}






\end{document}